\documentclass{aa}

\usepackage{graphicx}
\usepackage{txfonts}
\usepackage{color}

\begin{document}

   \title{A nuclear molecular outflow in the Seyfert galaxy NGC~3227}


   \author{A. Alonso-Herrero\inst{\ref{inst1}}
          \and
          S. Garc\'{\i}a-Burillo\inst{\ref{inst2}}
          \and
          M. Pereira-Santaella\inst{\ref{inst3}}
          \and
          R. I. Davies\inst{\ref{inst4}}
          \and
          F. Combes\inst{\ref{inst5}}
         \and
          M. Vestergaard\inst{\ref{inst6}, \ref{inst7}}
          \and 
          S. I. Raimundo\inst{\ref{inst6}}
          \and
          A. Bunker\inst{\ref{inst3}}
          \and
         T. D\'{\i}az-Santos\inst{\ref{inst8}}
         \and
          P. Gandhi\inst{\ref{inst9}}
          \and
          I. Garc\'{\i}a-Bernete\inst{\ref{inst10}}
          \and
          E. K. S. Hicks\inst{\ref{inst11}}
          \and
          S. F. H\"onig\inst{\ref{inst8}}
          \and
          L. K. Hunt\inst{\ref{inst12}}
          \and
          M. Imanishi\inst{\ref{inst13}, \ref{inst14}}
          \and
          T. Izumi\inst{\ref{inst13}}
          \and
          N. A. Levenson\inst{\ref{inst15}}
          \and 
          W. Maciejewski\inst{\ref{inst16}}
          \and
          C. Packham\inst{\ref{inst17}, \ref{inst13}}
          \and
          C. Ramos Almeida\inst{\ref{inst18}, \ref{inst19}}
          \and
          C. Ricci\inst{\ref{inst8}, \ref{inst20}}
          \and
          D. Rigopoulou\inst{\ref{inst3}}
          \and
          P. F. Roche\inst{\ref{inst3}}
          \and
          D. Rosario\inst{\ref{inst21}}
          \and
          M. Schartmann\inst{\ref{inst22}, \ref{inst4}}
          \and
          A. Usero\inst{\ref{inst2}}
          \and
        M. J. Ward\inst{\ref{inst21}}
}

   \institute{Centro de Astrobiolog\'{\i}a (CAB, CSIC-INTA), ESAC
     Campus, E-28692 Villanueva de la Ca\~nada, Madrid, Spain\\
              \email{aalonsol@cab.inta-csic.es}\label{inst1}
         \and
          Observatorio de Madrid, OAN-IGN, Alfonso XII, 3, E-28014 Madrid, Spain\label{inst2}   
          \and
        Department of Physics, University of Oxford, Keble Road, Oxford OX1 3RH, UK\label{inst3}   
        \and
        Max Planck Institut f\"ur extraterrestrische Physik Postfach 1312, D-85741 Garching bei M\"unchen, Germany\label{inst4}   
        \and
        LERMA, Obs. de Paris, PSL Research Univ., Coll\'ege de France,
        CNRS, Sorbonne Univ., UPMC, Paris, France\label{inst5}   
        \and
         DARK, The Niels Bohr Institute, University of Copenhagen,
         Vibenshuset, Lyngbyvej 2, DK-2100 Copenhagen O., Denmark\label{inst6}
         \and
         Steward Observatory, University of Arizona, 933 N. Cherry
         Avenue, Tucson, Arizona, USA\label{inst7}
         \and
         N\'ucleo de Astronom\'{\i}a de la Facultad de
         Ingenier\'{\i}a, Universidad Diego Portales, Av. Ej\'ercito
         Libertador 441, Santiago, Chile\label{inst8} 
         \and
         Department of Physics \& Astronomy, University of
         Southampton, Hampshire SO17 1BJ, Southampton, UK\label{inst9} 
         \and
         Instituto de F\'{\i}sica de Cantabria, CSIC-Universidad de
         Cantabria, E-39005 Santander, Spain\label{inst10} 
         \and
         Department of Physics \& Astronomy, University of Alaska Anchorage, AK 99508-4664, USA\label{inst11}
         \and
         INAF-Osservatorio Astrofisico di Arcetri, Largo E. Fermi 5, 50125 Firenze, Italy\label{inst12}
         \and
         National Astronomical Observatory of Japan, National
         Institutes of Natural Sciences (NINS), 2-21-1 Osawa, Mitaka,
         Tokyo 181–8588, Japan\label{inst13} 
         \and
         Department of Astronomical Science, SOKENDAI (The Graduate
         University of Advanced Studies), 2-21-1 Osawa, Mitaka, Tokyo
         181-8588, Japan\label{inst14} 
         \and
         Space Telescope Science Institute, 3700 San Martin Drive,
         Baltimore, MD 21218, USA\label{inst15}
         \and
         Astrophysics Research Institute, Liverpool John Moores
         University, IC2 Liverpool Science Park, 146 Brownlow Hill, L3
         5RF, UK \label{inst16}
         \and
         Department of Physics and Astronomy, University of Texas at
         San Antonio, 1 UTSA Circle, San Antonio, TX 78249,
         USA\label{inst17} 
         \and
         Instituto de Astrof\' isica de Canarias, Calle V\' ia
         L\'actea, s/n, E-38205, La Laguna, Tenerife,
         Spain\label{inst18}
         \and
         Departamento de Astrof\' isica, Universidad de La Laguna,
         E-38206, La Laguna, Tenerife, Spain\label{inst19}
         \and
         Kavli Institute for Astronomy and Astrophysics, Peking University, Beijing 100871, China\label{inst20}
         \and
         Centre for Extragalactic Astronomy, Durham University, South Road, Durham DH1 3LE, UK\label{inst21}
         \and
         Excellence Cluster Origins, Ludwig-Maximilians-Universität
         München, Boltzmannstr. 2,  85748 Garching, Germany\label{inst22}
}

   \date{Received:  --; accepted ---}

 
  \abstract
   {
ALMA observations have revealed nuclear dusty molecular
     disks/tori  with characteristic sizes 15-40\,pc in the few Seyferts and low
     luminosity AGN studied so far. 
     These structures are generally decoupled both morphologically and
   kinematically from the host galaxy disk. 
 We present 
ALMA observations of the CO(2--1) and CO(3--2)
     molecular gas transitions and associated (sub)-millimeter
     continua of  the nearby Seyfert 1.5 galaxy NGC~3227 
with angular resolutions $0.085-0.21\arcsec$ (7--15\,pc).
On large scales the cold molecular gas shows circular motions as
     well as streaming motions on scales of a few hundred
     parsecs associated with a large scale bar.
We fitted the nuclear ALMA 1.3\,mm emission with an unresolved
component and an extended component. The $850\,\mu$m emission 
shows at least two
     extended components, one along the major axis of the nuclear disk and the
     other along the axis of the ionization cone. The molecular gas in
     the central region  (1\arcsec$ \sim 73\,$pc)  
shows several CO clumps with complex kinematics which  appears to be
 dominated by non-circular motions. While we cannot demonstrate
 conclusively the presence of a warped nuclear disk, we also detected
 non-circular motions along the kinematic minor axis. They
 reach line-of-sight velocities of $v-v_{\rm sys} =150-200\,{\rm km\,s}^{-1}$.
Assuming that the radial motions are  in the plane of the galaxy, then we  interpret them as
a nuclear molecular outflow  due to molecular gas in the host galaxy being
entrained by the AGN wind. We derive molecular outflow rates of $5\,M_\odot\,{\rm
  yr}^{-1}$ and $0.6\,M_\odot\,{\rm
  yr}^{-1}$ at projected distances of
up to 30\,pc to the northeast and southwest of the AGN, respectively.
At the AGN location we estimate a mass in molecular gas of $5\times 10^{5}\,M_\odot$ 
and an equivalent average column density 
$N({\rm H}_2) = 2-3\times 10^{23}\,{\rm cm}^{-2}$ in the inner
15\,pc. The nuclear CO(2-1) and CO(3-2) molecular gas
and sub-mm continuum emission of NGC~3227 do
     not resemble the {\it classical} compact torus. Rather, these emissions
     extend for several tens of parsecs and appear
     connected with the circumnuclear ring in the host galaxy
     disk, as found in other local AGN.
}

   \keywords{ Galaxies: kinematics and dynamics --
                     Galaxies: Seyfert -- Galaxies: individual:
                     NGC~3227 -- Submillimeter: galaxies}

   \maketitle

\clearpage
%


\section{Introduction}
The detection of the hidden broad line region of NGC~1068 \citep{AntonucciMiller1985},  the
archetypical type 2  active galactic nucleus  (AGN), boosted the idea
of the so-called Unified Model. A dusty  molecular torus obscures the
view of the central engine in type 2 AGN along directions near the
equatorial plane while near polar directions allow the detection of the broad line
region thus classifying the AGN as a type 1. This torus was
proposed to be both geometrically and optically thick
\citep[e.g.,][]{Antonucci1993}. To explain the observed properties of NGC~1068
and other active galaxies,
the theoretical models initially favored a relatively compact torus 
\citep{PierKrolik1992,  Nenkova2002, Nenkova2008a,
  Nenkova2008b}. Subsequently, it was suggested 
that the torus is  part of a larger and more diffuse structure
extending for  up to 100\,pc \citep{PierKrolik1993, GranatoDanese1994,
  MaiolinoRieke1995, Efstathiou1995, Young1996}. However, only the
inner warm  parts of the
torus would be responsible for the near and mid-infrared emission \citep{Schartmann2008}.
Additionally, most local Seyfert galaxies with sufficiently high
signal-to-noise mid-infrared interferometric observations show
emission along the polar direction on parsec-scales 
 \citep{Tristram2009, Hoenig2013, LopezGonzaga2016}.
Theoretical arguments indicate that this phenomenon could be associated with an outflowing
torus, as already captured by dusty disk-wind models
\citep{ElitzurShlosman2006, HoenigKishimoto2017} and
radiation-driven outflow models \citep{Wada2012,  Wada2016}. Other alternatives
to the optically and geometrically thick torus, including nuclear warped
disks \citep[see][for Circinus]{Jud2017}
are discussed in detail by \cite{Lawrence2010}. See also
\cite{RamosAlmeida2017} for a recent review.

The excellent high angular resolution achieved with the Atacama Large
Millimeter Array  
(ALMA) is now revealing that relatively large (diameters of up to
20-50\,pc) molecular tori/disks 
appear to be  common in the few local Seyferts and low-luminosity AGN
observed so far \citep{GarciaBurillo2014,
  GarciaBurillo2016, Gallimore2016, Imanishi2018, AlonsoHerrero2018,
  Izumi2017, Izumi2018, Combes2019}. 
Moreover, these nuclear tori/disks appear to be decoupled both morphologically and kinematically
from their host galaxies. This is in good agreement with previous
results that found random orientations of the 
radio jets/ionization cones of local AGN with respect to their host galaxies
\citep[see e.g.,][]{Nagar1999, Kinney2000, Fischer2013}. Studying the mechanisms responsible for this
decoupling as well as the physical scales where they operate is thus
crucial for making progress in our understanding of the general
question of the fueling and nuclear obscuration of active galaxies. 

  This is the second paper of a series \citep[first paper,][]{AlonsoHerrero2018}
  based on several ALMA programs to observe the
cold molecular gas in the nuclear and circumnuclear regions of
an ultra-hard X-ray selected sample of nearby Seyfert galaxies. 
Our parent sample is drawn
from the X-ray {\it Swift}/BAT $14-195\,$keV all-sky 70 month catalog
\citep{Baumgartner2013}. Our main goal 
is to understand the connections between the cold and hot
molecular gas, the AGN torus, and nuclear/circumnuclear star
formation activity in local active galaxies. In this paper
we present a detailed study
of the cold molecular gas emission of the 
nearby Seyfert 1.5 galaxy NGC~3227. We use a distance of
$14.5\pm0.6$\,Mpc 
as derived from  AGN lags \citep{Yoshii2014} and thus
1\arcsec$\sim$73\,pc. The intrinsic $14-195\,$keV X-ray luminosity
is approximately $2\times 10^{42}\,$erg\,s$^{-1}$ \citep{Ricci2017}
for our assumed distance.
This galaxy is interacting with NGC~3226 and
presents evidence for large scale streaming motions detected in
neutral hydrogen \citep{Mundell1995_HI}. It is also located in a
group of 13-14 galaxies \citep{Davies2014}.  Finally, NGC~3227 is
  classified as SAB(s)a pec in the Third Reference Catalogue of Bright Galaxies
  \citep{deVaucouleurs1991}. However, \cite{Mulchaey1997} modeled  near-infrared
  observations of this galaxy, which revealed  the presence of a large
  scale bar.

  \begin{figure*}
   \centering
  \includegraphics[width=18cm]{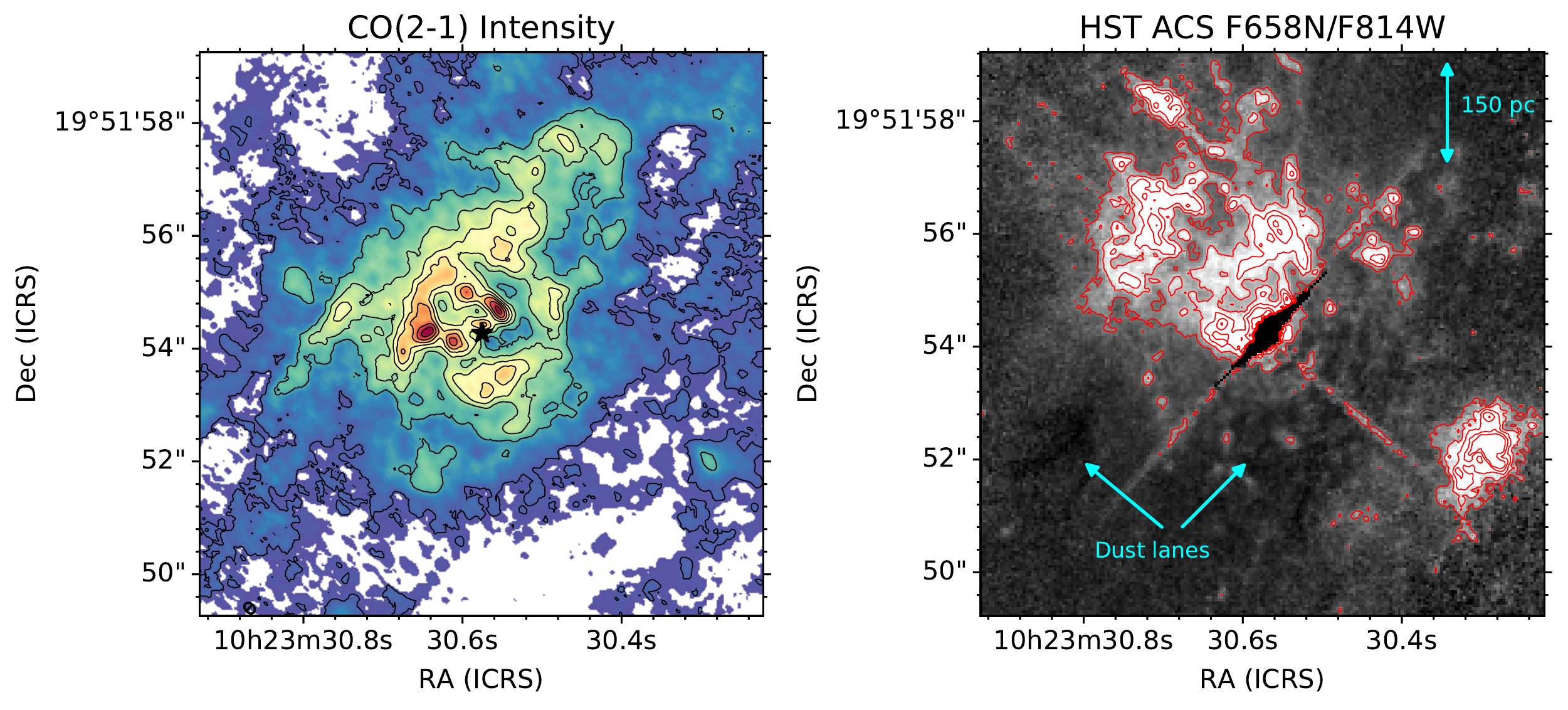}
   \caption{{\it Left panel}. ALMA CO(2--1) integrated molecular line
     map (0$^{\rm th}$-order moment map)  created using  a $3\sigma$ threshold. The image displays 
the central $10\arcsec \times 10\arcsec$ ($730\,{\rm
  pc} \times 730\,{\rm pc}$) and was created
     from the $b=0.5$ weight data cube. We show the image and contours
     with a linear scale. The CO(2--1) contours are 0.0925, 0.515,
     0.9375,  1.36,    1.7825,  2.205,   2.6275 and 3.05\,
     Jy\,km\,s$^{-1}$  beam$^{-1}$. The star symbol indicates the AGN location, based on
     the peak of the unresolved 1.3\,mm continuum emission (see text). 
{\it Right panel}. HST/ACS
     pseudo  H$\alpha$+[\ion{N}{ii}] line map (highlighted in light colors) plus
     extinction map (highlighted in dark colors)
     constructed as an F658N/F814W map (in arbitrary units) and showing the same FoV as
the ALMA CO(2-1) image. Note that the broad-band
F814W image is saturated at the nuclear position. }
              \label{fig:ALMAlargeFoV}%
    \end{figure*}


\cite{Schinnerer2000} used the Plateau de
Bure interferometer (PdBI) to obtain CO(1--0) and CO(2--1) observations of
NGC~3227 with angular
resolutions in the $\sim 0.5-1.5\arcsec$ range. They
detected apparent counterrotation in the nuclear region
of this galaxy and interpreted this kinematic signature as evidence of the warping
of a thin inner molecular disk starting at an approximately
outer radius of 1\arcsec. \cite{Davies2006}, on the other hand, based
on the high velocity dispersion of the hot molecular gas traced by the
H$_2$ $2.12\,\mu$m 1--0 S(1) emission line, argued that
the nuclear kinematics could be explained with a nuclear thick disk with turbulent
motions. In this paper we present ALMA Cycle 4 observations of the
molecular gas CO(2--1) and CO(3--2) transitions and associated
continuum of the nuclear region of NGC~3227, 
obtained with angular resolutions
ranging from 0.21\arcsec \, to 0.085\arcsec. The paper is organized as follows.
Section~\ref{sec:observations} presents the ALMA observations and
ancillary data used in this paper.  In Sections~\ref{sec:continuum}
and \ref{sec:coldgas} we discuss the (sub)-millimeter continuum and
molecular gas CO(2--1) and CO(3--2) properties.
We model the CO(2-1) molecular gas kinematics in Section~\ref{sec:barolo}.
In Section~\ref{sec:molgascolden} we derive the cold molecular gas
mass and column density in the nuclear region of NGC~3227.
Finally in Section~\ref{sec:conclusions} we present our conclusions.

\section{Observations and Data Reduction}\label{sec:observations}

\subsection{ALMA Band 6 and Band 7 observations}

We obtained Band 6 ALMA observations of NGC~3227 between March and 
September 2017 using the 12\,m array in two configurations (compact
with baselines between 15 and 460 m and extended with baselines
between 20 and 3700 m) and the Atacama Compact Array (ACA) with
baselines between 9 and 50 m. These observations were part of the
project 2016.1.00254.S (PI: A. Alonso-Herrero). 
We defined two spectral windows of 1.875\,GHz bandwidth (with 3.9\,MHz
$\sim$ 5\,km
s$^{-1}$ channels), at the observed frequency of the CO(2-1)
transition (229.8\,GHz) and the other at an observed frequency of
$\sim$231\,GHz (1.3\,mm) to measure the sub-millimeter continuum. 
We also used Band 7 ALMA data of NGC 3227 observed between August
and September 2017 with the 12\,m array in a single configuration with
baselines between 21 and 3700 m observed for the 2016.1.01236.S (PI:
M. Vestergaard)
project. A spectral window was centered on the CO(3-2) transition
(344.4\,GHz) and other three spectral windows at 342.6, 354.6, and
356.5\,GHz to measure the continuum (approximately $850\,\mu$m).

\begin{table}
\caption{Details of the ALMA continuum observations}             
\label{tab:ALMAcontinuum}      
\centering                          
\begin{tabular}{c c c c c}        
\hline\hline                 
ALMA & weight & Beam size & PA$_{\rm beam}$ & rms \\    
          &             & ($\arcsec \times \arcsec$) & (\degr) & (mJy beam$^{-1}$)\\
\hline                        
   Band 6 & $b=0.5$   & $0.196\times0.151$& 48.1 & $3.9\times 10^{-2}$\\      
   Band 6 & $b=-0.5$ & $0.175\times0.127$ & 45.7 & $7.7\times 10^{-2}$\\
   Band 7 & natural     & $0.094\times0.085$ & 16.6 & $2.0\times 10^{-2}$\\
\hline                                   
\end{tabular}
\end{table}

We calibrated and imaged the data using the ALMA reduction software CASA 
\citep[v.5.1, ][]{McMullin2007}. 
For the 12\,m array data (Band 6 and 7), either
J0854+2006 or J1058+0133 were used for both the bandpass and amplitude
calibrations and J1025+1253 was used for the phase calibration. For
the Band 6 ACA data, J1058+0133 was used for the bandpass calibration,
Ganymede for the flux calibration, and J1041+0610 for the phase
calibration. The Ganymede flux was estimated using the
Butler-JPL-Horizons 2012 model.  The consistency of the flux calibration was verified by comparing the 
observed amplitudes as a function of the uv distance for all the 
configurations. Since the amplitudes of the overlapping 
baselines are similar we can conclude that the flux calibration is 
consistent. The J0854 and J1058 fluxes are 2.5 and 3.1\,Jy at
229\,GHz.  Finally, we note that we used the same phase calibrator for 
 band 6 and 7 except for the ACA band 6 data (J1041+0610). However, the
ACA data should not significantly affect the emission detected on the scales 
discussed in this work and therefore, all the data are referenced to the 
same phase calibrator.

 \begin{table*}
\caption{Details of the ALMA CO(2-1) and CO(3-2) observations}             
\label{tab:ALMAmoleculargas}      
\centering                          
\begin{tabular}{c c c c c c c}        
\hline\hline                 
ALMA & Transition & weight & Beam size & PA$_{\rm beam}$ & rms per
                                                           channel &
                                                                     vel res.\\    
          &   &          & ($\arcsec \times \arcsec$) & (\degr) & (mJy km s$^{-1}$ beam$^{-1}$) &
(km s$^{-1}$)\\
\hline                        
Band 6 &   CO(2--1) & $b=0.5$ & $0.214\times0.161$ & 42.0& $4.8\times
                                                    10^{-1}$ & 15 \\
Band 7 &   CO(3--2) & natural    & $0.095\times0.085$ & 21.3 &
                                                               $5.1\times
                                                               10^{-1}$
                                                                   & 10\\
\hline                                   
\end{tabular}
\end{table*}

We subtracted the corresponding continua from the CO(2-1) and CO(3-2)
spectral windows directly
in the visibility data by fitting the continuum emission with a
constant in the line-free channels. 
Then, we combined and cleaned the data using the CASA 
{\sc clean} task.
The output frequency reference frame was set to the kinematic local standard of rest (LSRK).
For the Band 6 continuum images, we used a  continuum spectral window
which we cleaned with the CASA {\sc clean} task using the Briggs weighting
\citep{Briggs1995}, 
with robustness parameters of $b=0.5$ and
$b=-0.5$. These result in a lower angular resolution and 
higher sensitivity,  and higher angular resolution  and slightly
decreased sensitivity, respectively. We summarize the beam
properties and sensitivities of the Band 
6 and Band 7 continuum maps in Table~\ref{tab:ALMAcontinuum}. 
For the Band 7 data, we combined the line-free channels of the 4
spectral windows also using the CASA {\sc clean} task and a  natural
weight (b=2). 
For the CO(2--1) and CO(3--2) transitions we produced cleaned data cubes
with weights $b=0.5$ and natural, respectively and the resulting beam
sizes and sensitivities are listed in Table~\ref{tab:ALMAmoleculargas}.
As can be seen from these two tables, our ALMA observations have
angular resolutions between 0.085 and 0.21\arcsec. For the assumed
distance to NGC~3227 they probe physical resolutions in the 7-15\,pc
range.

From the Band 6 and Band 7 data cubes we produced maps of the CO(2--1) 
and CO(3--2)  integrated intensity,
mean-velocity field, and velocity dispersion, respectively. For this
we used the GILDAS\footnote{http://www.iram.fr/IRAMFR/GILDAS}
{\sc moment} task and set the threshold limit to 
$3\sigma$ and $5\sigma$. In Figure~\ref{fig:ALMAlargeFoV}  (left panel), we show the integrated CO(2–1)
molecular line map created from the
Band 6 data cube with $b=0.5$ and the $3\sigma$ detection threshold. We show an
approximate  field of view (FoV)  of $10\arcsec \times
10\arcsec$, which corresponds to a circumnuclear region of $730\,{\rm
  pc} \times 730\,{\rm pc}$ in size. In this figure and following figures, the coordinates
   are in the International Coordinate Reference System (ICRS) frame.

\subsection{Ancillary archival HST images}
We downloaded from the Hubble Legacy Archive (HLA) images taken with
the Advanced Camera for Surveys (ACS) on board of the  {\it Hubble Space Telescope}
(HST) using the narrow-band filter F658N, which contains the
H$\alpha$+[\ion{N}{ii}] emission lines, and the broad-band filter
F814W (proposal ID: 9293). The HLA images are fully reduced and
drizzled to a pixel size of 0.05\arcsec. We 
adjusted the HST/ACS astrometry for NGC~3227 by fixing the coordinates of the bright nuclear  point
source to that of the peak of the unresolved ALMA band 6 continuum at 1.3\,mm
(see Section~\ref{subsec:continuum_morphology}). To trace the emission
from the NGC~3227 ionization cone and circumnuclear \ion{H}{ii} regions
\citep[see][]{GonzalezDelgado1997}  at high angular resolution, we
constructed a pseudo H$\alpha$+[\ion{N}{ii}]  line map by dividing the F658N image by that of the nearby
broad-band continuum traced by the F814W image (Figure~\ref{fig:ALMAlargeFoV}, right panel). This map also
highlights the regions with high extinction. These are seen as dust
lanes toward the west, southwest and southeast of the nucleus (marked
in the figure), which
are due mostly to material in the near side of the
galaxy \citep[see][]{Martini2003, Davies2014}.

In Figure~\ref{fig:ALMAlargeFoV}  we show the F658N/F814W map for
the same FoV as the ALMA CO(2--1) integrated line emission map.
As inferred from optical imaging, line ratios and emission line
kinematics  of this galaxy \citep{Mundell1995, GonzalezDelgado1997},
the line emission to the north, northwest and 
northeast is mostly photoionized by the AGN and it is likely tracing
mostly one side of the ionization cone of this galaxy
\citep{Fischer2013}, whereas the emission at $\sim 4\arcsec$ \, to the southwest of the AGN
is mostly due to star formation activity in the disk of the
galaxy. Further indication for AGN photoionization to the north and
northeast is the detection of
outflowing ionized gas at PA$\sim 10\degr$, reaching velocities of up
to $900\,{\rm km\,s}^{-1}$ \citep[see][]{Barbosa2009}. Moreover, modelling of the ionized
gas kinematics finds a position angle (measured east from the north) for the ionization cone of
PA$_{\rm cone} = 30\degr$ \citep{Fischer2013}. This value is also in
good agreement with the value derived from optical polarization,
assuming that the polarization is due to polar scattering \citep{Smith2004}.

\begin{figure*}
\centering
\includegraphics[width=19cm]{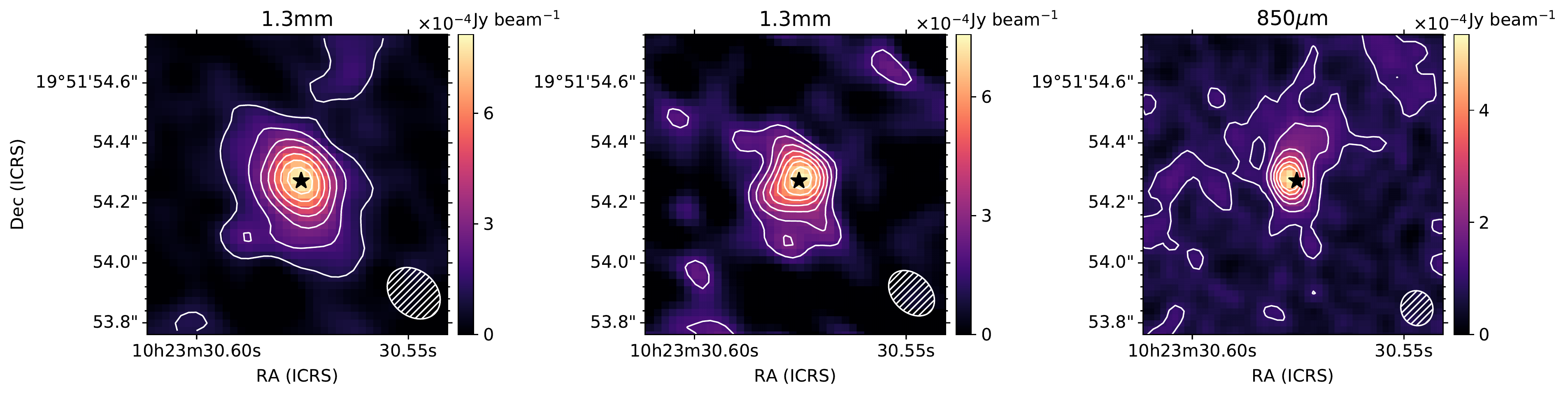}
\caption{ALMA continuum images in Band 6 at $\sim 1.3\,$mm from the
     $b=0.5$ weight data cube (left) and $b=-0.5$ weight data cube
     (middle) and in Band 7 at $\sim 850\,\mu$m from the natural weight data cube
     (right). The FoV is $1\arcsec \times 1\arcsec$ which corresponds
     to approximately the central $73\,{\rm pc } \times 73\,{\rm pc}$
     region. In all images the first contour corresponds to the
     $2.5\sigma$ level and the color bars indicate the flux in units
     of $10^{-4}\,$Jy beam$^{-1}$. The hatched ellipses show the corresponding
     beam size and orientation in each image (see
     Table~\ref{tab:ALMAcontinuum}) and the star symbol indicates the
     approximate location of the peak of the unresolved 1.3mm continuum emission (see text).}
              \label{fig:ALMAcontinua}%
\end{figure*}

\section{Nuclear continuum emission}\label{sec:continuum}

\subsection{Morphology}\label{subsec:continuum_morphology}
In Figure~\ref{fig:ALMAcontinua} we show the images of the 1.3\,mm and
$850\,\mu$m continua of NGC~3227 for the nuclear $1\arcsec \times
1\arcsec$ region ($73\,{\rm pc } \times 73\,{\rm pc}$). The continuum
at both wavelengths appears clearly extended. The 1.3\,mm continuum
presents a complex morphology with an unresolved source and probably
several extended components (see below). The 1.3\,mm continuum map derived from
the $b=0.5$ weight data cube shows an extended component along the
northeast-southwest direction, that is, at a PA$\simeq 30-40\degr$. This direction
coincides  with 
the  axis of the projected H$\alpha$+[\ion{N}{ii}]  emission (see the
right panel of Figure~\ref{fig:ALMAlargeFoV}),  
which traces mostly emission on one side of the cone in the ionization cone \citep[see
e.g.][]{Mundell1995, GonzalezDelgado1997, Barbosa2009, Fischer2013}. The 1.3\,mm
continuum map derived with the $b=-0.5$ weight, which has higher angular
resolution (see Table~\ref{tab:ALMAcontinuum} and middle panel of
Figure~\ref{fig:ALMAcontinua}), shows more clearly that in addition to the emission in
the 
direction of the ionization cone, there is another component nearly
perpendicular to it (see below). 

For simplicity we modelled the 1.3\,mm continuum in the {\it uv} plane using
only two components, namely a point
source and a Gaussian source  to encompass all the extended emission. We left the positions of the two
  sources free and we only fixed the width of the extended component
  to FWHM=$0.35\arcsec$ due to the limited signal-to-noise ratio of
  this component. For the
unresolved component we obtained the following coordinates (ICRS frame)
RA(ICRS)=10$^{\rm h}$23$^{\rm m}$30.574$^{\rm s}$ and
Dec(ICRS)=+19$\degr$51\arcmin54.278\arcsec. These coordinates are,
within the errors, coincident with source C detected at 1.7
and 5\,GHz by
\cite{Bontempi2012} using high angular resolution European VLBI Network (EVN) radio
observations. Given the high brightness temperature and spectral index
of source C, these authors identified it as the position of the AGN of
NGC~3227. The 1.3\,mm extended component does 
not peak exactly at the same position but it appears offset by
approximately 0.07\arcsec \, to the southeast of the AGN position.
This offset appears to be real as it is a factor of a few above the
positional uncertainties of the phase calibrator used
during the data reduction. 
The derived fluxes at 1.3\,mm are $0.51\pm 0.15\,$mJy and
$1.92\pm0.21\,$mJy for the unresolved and resolved components, respectively.

\begin{figure}
\centering
\includegraphics[width=9cm]{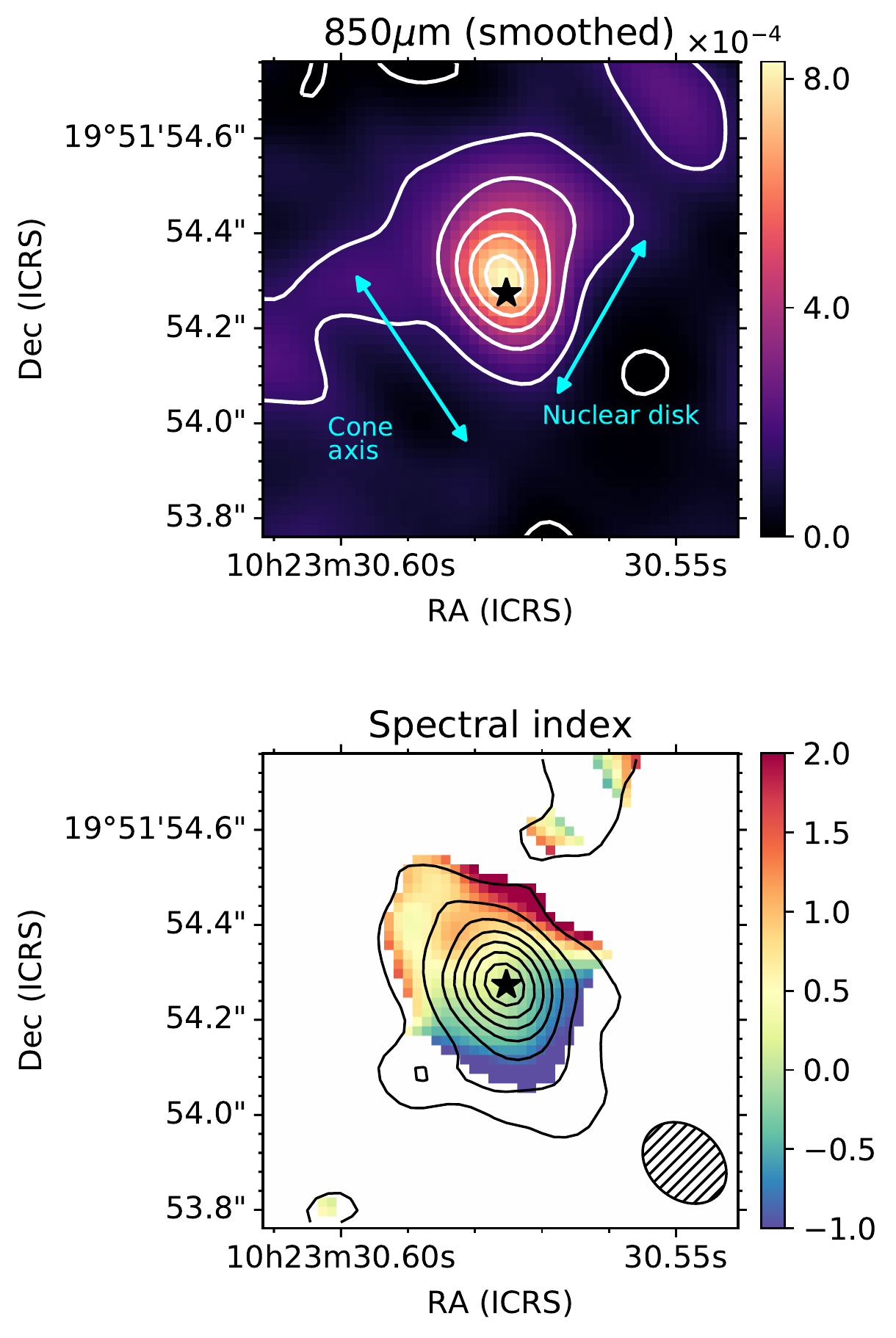}

\caption{{\it Top panel.} In color and  white contours
  is the map of the $850\,\mu$m continuum emission (in arbitrary units)
degraded to the beam size and orientation of the 1.3mm continuum map with $b=0.5$
weight (see Table~\ref{tab:ALMAcontinuum}). 
This was necessary to construct the spectral index map. The map is shown for the nuclear
$1\arcsec \times 1\arcsec$  region. We mark the approximate
orientations of the ionization cone axis and the nuclear star forming disk.
{\it Bottom panel.} In color is the map of the spectral  index
$\alpha$  as
derived from the ALMA Band 6 and  Band 7 continuum observations using
only pixels detected at $>2.5\sigma$. The FoV is as the top
panel. The map has the same angular
resolution as the Band 6 map with the $b=0.5$ weight (see
Table~\ref{tab:ALMAcontinuum}). The color bar indicates the values of
$\alpha$ (see text). The contours are the Band 6 continuum emission
at 1.3\,mm as in the left panel of 
Figure~\ref{fig:ALMAcontinua} and the star symbol marks the AGN position.}\label{fig:ALMAspectralindex}%
\end{figure}

The $850\,\mu$m continuum map (Figure~\ref{fig:ALMAcontinua}, right
panel)  shows that most of the nuclear emission at this wavelength
has a disk-like morphology extending for about 0.6\arcsec \,
($\sim 45\,$pc) and oriented at PA$\sim -20\degr$ to
$-30\degr$, which is  perpendicular to the axis of the ionization
cone. There is also fainter emission
in the direction of the ionization cone (see also the top panel of
Figure~\ref{fig:ALMAspectralindex}). The $850\,\mu$m
continuum morphology shows a remarkable resemblance to the SINFONI 
Br$\gamma$ narrow component map presented by
\cite{Davies2006} (their figure~3, with a similar FoV as our continuum
maps). These authors interpreted this  Br$\gamma$ emission
as originating mostly in star formation activity and is
probably associated with the nuclear stellar disk. 
There is also evidence for the presence of nuclear on-going/recent
star formation activity from the detection of polycyclic aromatic
hydrocarbon emission on scales of $\sim 1\arcsec-0.5\arcsec$ 
\citep{Imanishi2002, Rodriguez-Ardila2003, Davies2007, Esquej2014,
  AlonsoHerrero2016}. 
Moreover, \cite{Davies2006} also
suggested that the MERLIN 6\,cm radio continuum 
emission \citep{Mundell1995}, which has an  extended component approximately in the same
direction as the nuclear disk, might be due to the superposition of many supernova
remnants rather than due to a radio jet  \citep[see also][]{Chapman2000}.

 \begin{figure}
   \centering
  \includegraphics[width=10cm]{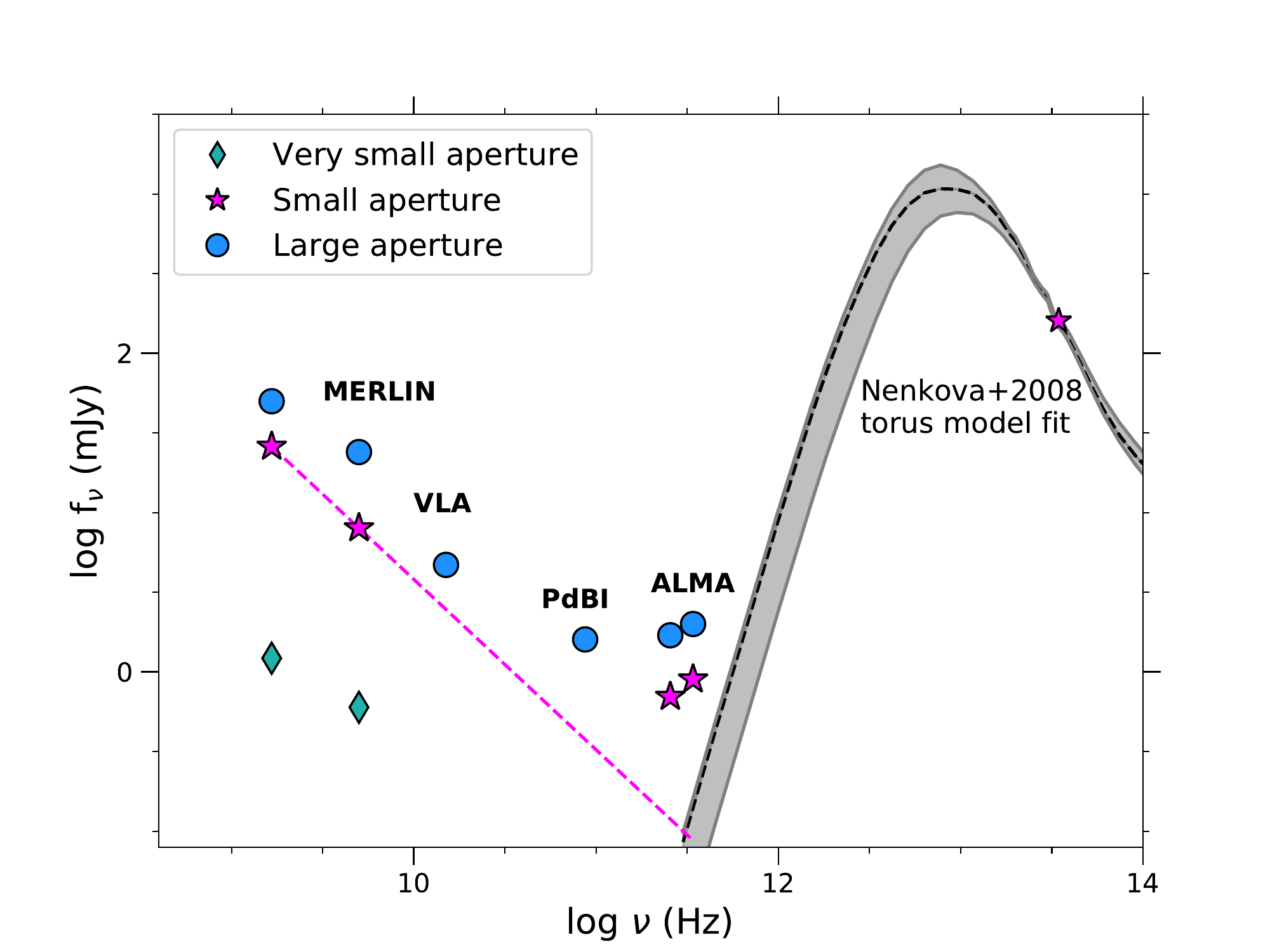}
   \caption{Nuclear continuum SED from cm wavelengths to the infrared of
NGC~3227. The {\it small} aperture ($\sim 0.1-0.2\arcsec$) MERLIN data points
are for the southern source of \cite{Mundell1995}. The {\it large}
aperture ($>0.2\arcsec$ \, to $\le1\arcsec$) 
VLA and PdBI data points are from \cite{Nagar2005} and \cite{Sani2012},
respectively. All the  ALMA continuum data points are from this
work. The {\it very small} aperture (a few milli-arcseconds) points are the EVN radio fluxes from
\cite{Bontempi2012}. The fit (median dashed black line, with the $\pm 1\sigma$ uncertainty
plotted with the shaded grey region)  to the 
near and mid-infrared unresolved emission. using the
\cite{Nenkova2008a, Nenkova2008b} clumpy torus models is from
\cite{GarciaBernete2019}. We also plotted the observed
mid-infrared data point for reference although the fitted data
also included  near-infrared photometry and mid-infrared
spectroscopy. The magenta dashed line is the fit to
the MERLIN radio points and extrapolated to the ALMA wavelengths.}
              \label{fig:SED}%
    \end{figure}

\subsection{Spectral index}\label{subsec:spectralindex}
To investigate further the nuclear  (sub)-millimeter continuum
emission of NGC~3227, we constructed a map of the 
 1.3\,mm to $850\,\mu$m spectral index, 
defined as $f_\nu \propto \nu^\alpha$,   with the ALMA Band 6 and
  Band 7 
continuum images. We used GILDAS to
convolve the $850\,\mu$m continuum map to the beam size and
orientation of the  1.3\,mm continuum map for the $b=0.5$ weight
($0.196\arcsec \times0.151\arcsec$ at PA$_{\rm beam}=48\degr$). We
show the result in the bottom panel of Figure~\ref{fig:ALMAspectralindex}. For the
spectral index map we only used pixels where both continua were
detected at a $2.5\sigma$ level or higher. With the adopted definition
of the spectral index, positive values in the range  $\alpha=+1$ to
$\alpha= +2$
would indicate a thermal dust component (from the dusty molecular
torus or dust heated by star formation or both). 
Pure synchrotron radiation (from a jet or supernova remnants)  has
$\alpha=-0.8$ (for non-beamed emission), and thermal free-free
emission (associated for instance with \ion{H}{ii} regions) takes values
of $\alpha \simeq 0$ for optically thin emission 
and $\alpha \simeq +2$ for optically thick emission. More complicated
scenarios, which have been discussed for NGC~1068, include synchrotron
emission with free-free absorption and electron scattered synchrotron
emission \citep{Hoenig2008, Krips2011, Pasetto2019}. These processes result in  positive spectral indices.
Finally optically-thick synchrotron with $\alpha \sim 0$  can also be
relevant to low-luminosity AGN \citep{Koljonen2015}. 

 \begin{table*}
\caption{Nuclear measurements.}             
\label{tab:nuclear}      
\centering                          
\begin{tabular}{c c c c c c c c}        
\hline\hline                 
Aperture & \multicolumn{2}{c}{Flux density} & \multicolumn{2}{c}{Line
                                              flux} & Mass \\
  & 1.3\,mm & $850\,\mu$m & CO(2--1) & CO(3--2) & Mol Gas\\
($\arcsec \times \arcsec$) & (mJy) & (mJy) & (Jy km\,s$^{-1}$)& (Jy km\,s$^{-1}$) & ($M_\odot$) \\
  \hline                        
  $0.2 \times 0.2$ & 0.7 & 0.9 & 1.2 & 1.7 & $5\times 10^5$\\
  $0.5 \times 0.5$ & 1.7 & 2.0 & 7.3 & 14.3 & $3\times 10^6$\\
  \hline                                   
\end{tabular}

Notes.--- The two square apertures are referred to as {\it small} and {\it
  large} in Figure~\ref{fig:SED} (see also text). The molecular gas
masses are estimated in Section~\ref{sec:molgascolden}.
\end{table*}

\begin{figure*}
   \centering

\vspace{-0.5cm}  
\includegraphics[width=8.cm]{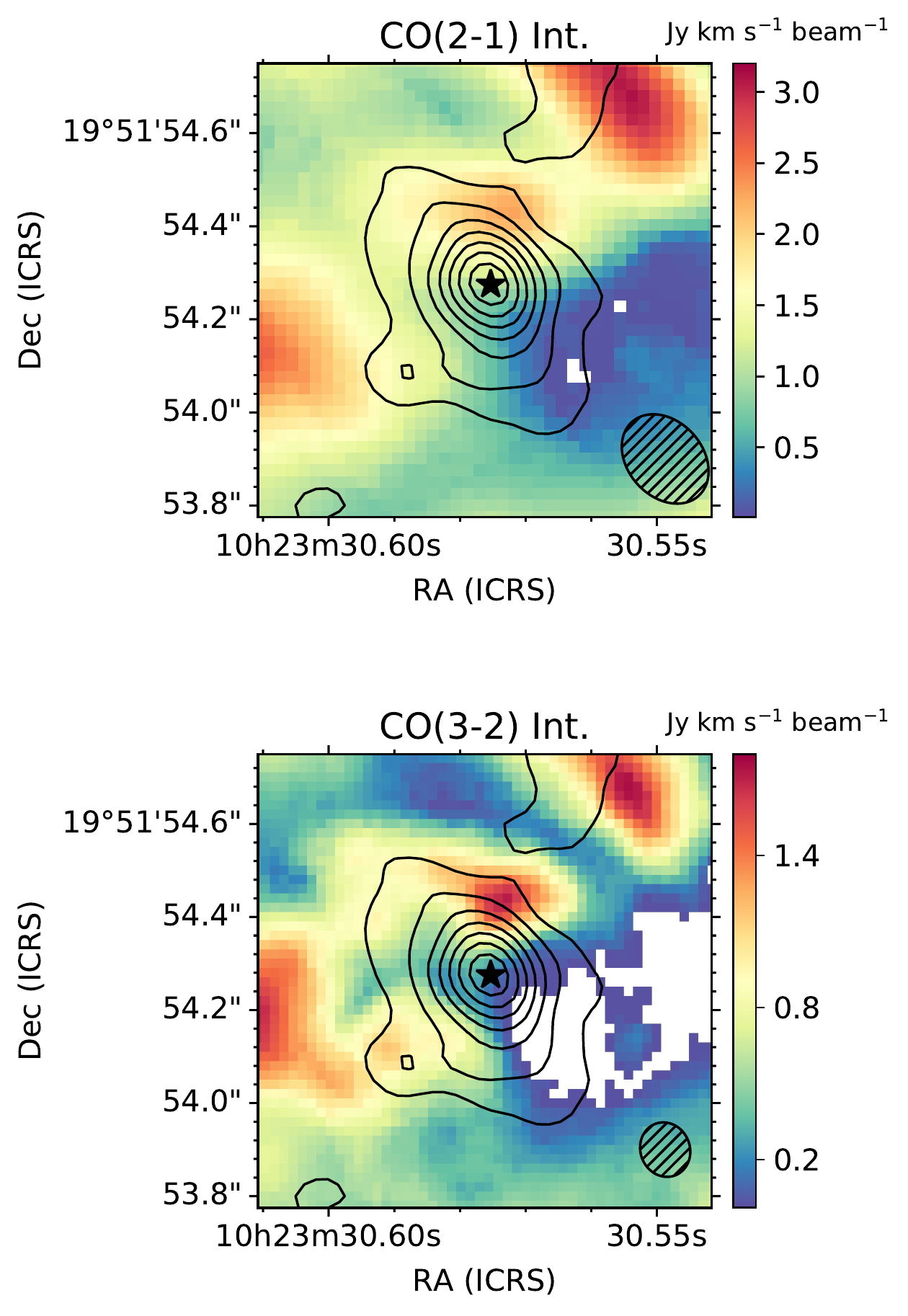}
\includegraphics[width=8.cm]{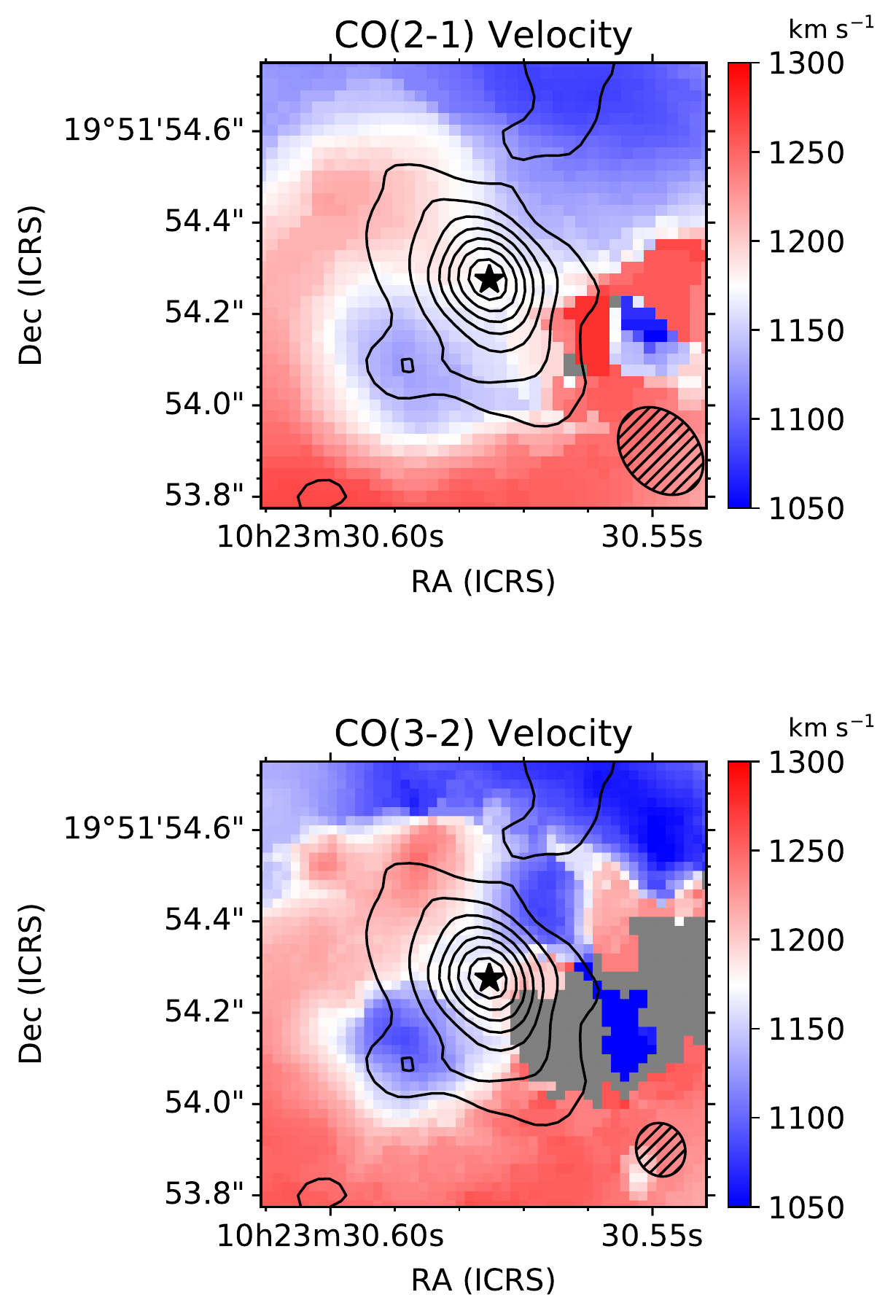}

  \vspace{-0.3cm}
  \caption{Zoom-in on to the CO(2--1) and CO(3--2)  emission in the nuclear
     $1\arcsec \times 1\arcsec$ region, in the top and  bottom panels
     respectively. These moment maps were created 
     with the GILDAS {\sc moment} task using a $5\sigma$ detection threshold and show
     the  integrated emission (0$^{\rm th}$-order moment map) in the left
     panels and the mean velocity field (1$^{\rm st}$-order moment map) in the
     right panels.  The hatched ellipses show the corresponding
     beam size and orientation of each image (see
     Table~\ref{tab:ALMAmoleculargas}).
     The contours in all panels are the ALMA   continuum emission at 1.3\,mm as in the left panel of
   Figure~\ref{fig:ALMAcontinua} and the star symbol marks the AGN position. }
              \label{fig:ALMAsmallCOmaps}%
    \end{figure*}

\begin{figure*}
   \centering
  \includegraphics[width=16.5cm]{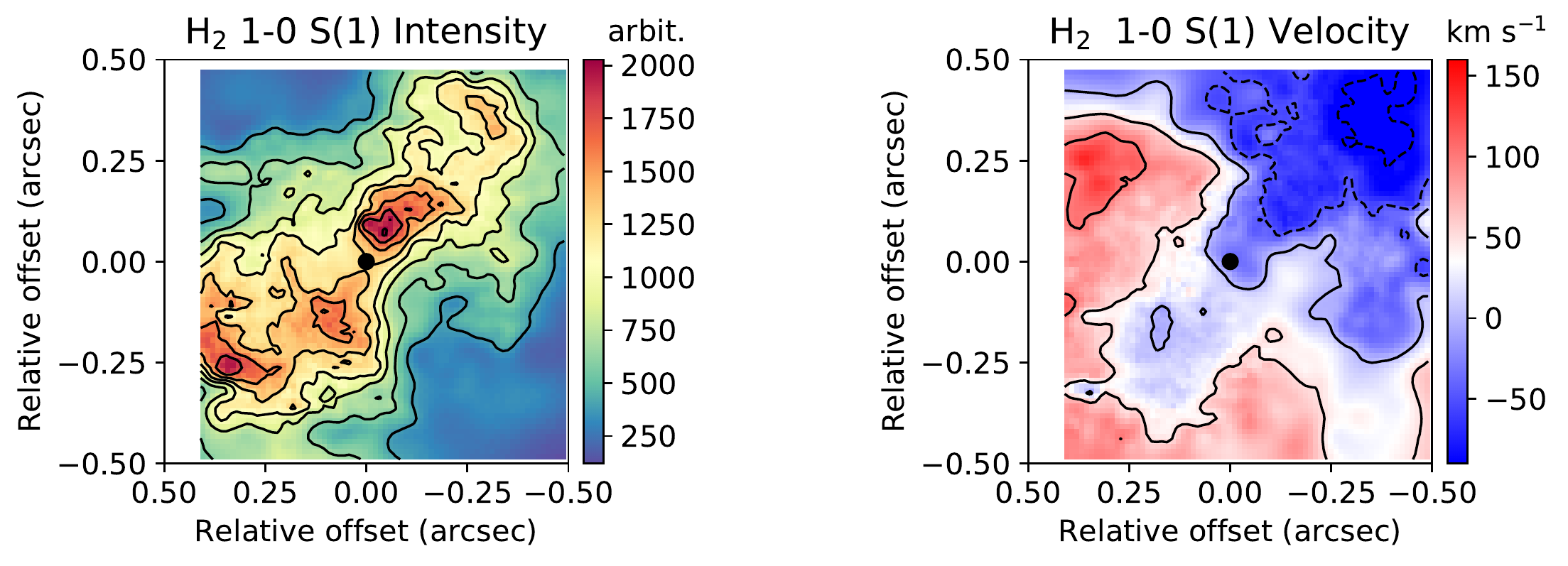}
  \vspace{-0.3cm}
   \caption{VLT/SINFONI H$_2$ 1-0 S(1) intensity (left) and velocity (right) maps from \cite{Davies2006} showing a FoV similar to that of
    Figure~\ref{fig:ALMAsmallCOmaps}. The maps were derived by fitting a Gaussian to the emission
     line. The black dot marks the approximate
     location of the peak of the near-infrared continuum. The
     velocities are relative to this peak. To
     facilitate the comparison with the CO(2--1) and CO(3-2) emission, we show the same velocity range as the
     right panels of 
 Figure~\ref{fig:ALMAsmallCOmaps}.}
              \label{fig:SINFONI}%
    \end{figure*}

In Figure~\ref{fig:ALMAspectralindex} (bottom panel) we show the spectral index
map of the nuclear
region of NGC~3227. The spectral index map displays a range of values. These variations
are 
real since they take place  on scales larger than the map beam. This
suggests that the (sub)-millimeter emission in this region is a  combination of different
mechanisms. To the south and southwest of the AGN position out to projected distances of
0.2\arcsec, the spectral index
appears to be mostly negative. This might indicate that synchrotron radiation
dominates in this region. This direction coincides with that
of the south part of the cone, which is  obscured in the optical
due to extinction from the host galaxy\footnote{The near side of the galaxy is to
the southwest  assuming that the spiral arms are trailing.}, as inferred from the modeling of the
optical
ionized gas kinematics \citep{Fischer2013}. 

To the north and northeast of the AGN, 
the observed spectral index could reflect a combination of thermal
processes and synchrotron emission. One possible interpretation 
for the extended 1.3\,mm emission at PA$\sim 30\degr$ to the northeast would be a
combination of polar dust emission and emission produced in a radio jet. In the
mid-infrared this galaxy shows  extended
$8-12\,\mu$m emission towards the north, northeast and south  of the AGN \citep[scales of
approximately 2\arcsec, see][]{AlonsoHerrero2016, Asmus2016,
  GarciaBernete2016}, which might be  tracing dust emission in the
ionization cone. If we interpret the non-thermal part of the extended  1.3\,mm emission as
a radio jet, then there would not be a large misalignment with the ionization cone emission and radio. It is
worth mentioning that \cite{Mundell1995} noted that there is also
18\,cm
radio emission detected perpendicular to PA$_{\rm radio}=-10\degr$, which could
be associated with the ALMA extended emission at 1.3\,mm. This extended
radio emission at a position angle similar to that of the ionization
cone can also be seen  at  6\,cm \citep[see figure~9 of][]{Davies2006}.

Along the nuclear disk at PA$\sim -20\degr$ and more importantly towards the northwest 
of the AGN the spectral
index takes positive values. This could be associated
with the on-going or recent star formation activity in the nuclear disk
\citep{Davies2006, Davies2007} via thermal processes such  as dust emission or free-free emission due to H\,{\sc ii} regions or both.

\subsection{Nuclear radio to sub-millimeter SED}
In addition to   the ALMA spectral index map, we
also compiled radio to (sub)-millimeter small aperture measurements to
construct the nuclear spectral energy distribution (SED) shown in
Figure~\ref{fig:SED}. For the radio
part we used the small aperture ($\sim 0.1-0.2\arcsec$) MERLIN 18\,cm and
6\,cm data points
of the southern source of \cite{Mundell1995}, which we associate
with the AGN position. From the ALMA 1.3\,mm and $850\,\mu$m continuum maps,
we measured the flux densities through a square aperture 0.2\arcsec
\, ($\sim 15\,$pc) on
a side. We also measured the ALMA continuum flux densities for a
larger square aperture  0.5\arcsec \, on
a side to approximately encompass the emission from the nuclear disk
detected at $850\,\mu$m (see Figure~\ref{fig:ALMAcontinua}).
Table~\ref{tab:nuclear} lists the ALMA continuum measurements. 
We compare them with the Very Large Array (VLA) 5\,GHz total flux
density of \cite{Nagar2005} and the 3\,mm flux density measured within
approximately 1\arcsec \, using the PdBI by \cite{Sani2012}.
For the identification of the AGN we show the EVN radio flux density
for source C of \cite{Bontempi2012}, which has a size of a few milli-arcseconds. We also
show the fit to the nuclear infrared emission of NGC~3227 (typical angular resolutions
$0.2-0.3\arcsec$) using the \cite{Nenkova2008a,
  Nenkova2008b} clumpy torus models derived by \cite{GarciaBernete2019}.

The derived spectral index fitted to the MERLIN nuclear values,
$\alpha_{\rm MERLIN}=-1.1$, is similar \citep[see
also][]{Mundell1995} to that
derived between 6 and 20\,cm by   \cite{Edelson1987} for NGC~3227 and
other Seyfert galaxies. The simplest interpretation is that it
reflects mostly synchrotron emission at these frequencies.
If we extrapolate this fit to the ALMA Band 6 and Band 7 frequencies (dashed
magenta line in Figure~\ref{fig:SED}), we can see that the contribution from
optically-thin synchrotron emission to the ALMA bands would be
small. However, the core emission could be self-absorbed. It is also 
possible that the low frequency spectral index of $-1.1$ may not hold all the way to the
mm wavelengths \citep{Behar2018}. 
Similarly, the predicted emission from the fitted \cite{Nenkova2008a,
  Nenkova2008b} clumpy torus model
at 1.3\,mm and $850\,\mu$m would only account for a small
fraction of the observed emission at these frequencies, even for the
0.2\arcsec\, aperture.  This suggests that even at $850\,\mu$m within
the inner $\sim 14\,$pc of NGC~3227, there
is a significant contribution from star formation likely giving rise to
thermal and non-thermal emission in this region. Indeed, \cite{Davies2006}
estimated that in this nuclear region approximately 75\% of the narrow
component Br$\gamma$ emission might originate in star formation
\citep[see also ][]{Davies2007}. Alternatively there could be a more
extended cold disk component, still heated by the AGN, which is not
captured by the clumpy torus models.

\begin{figure*}
   \centering
  \includegraphics[width=19cm]{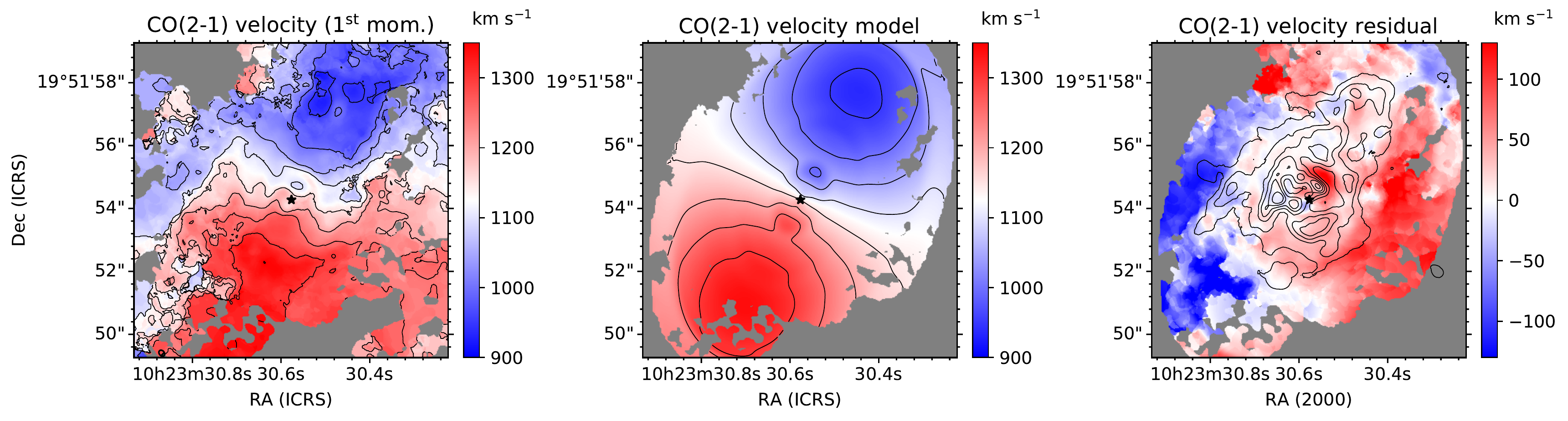}
   \caption{$^{\rm 3D}$BAROLO   maps of the observed CO(2--1) mean velocity field (1$^{\rm st}$-order 
     moment, left panel), the fitted
     model of a rotating
     disk with fixed PA, inclination (see Table~\ref{tab:BAROLOmodels}
     and text), systemic velocity and kinematic center 
     (middle panel), and the
     residual mean-velocity field (right panel).  The FoV is $10\arcsec \times 10\arcsec$, as in
     Figure~\ref{fig:ALMAlargeFoV}. The color bars indicate
     line-of-sight velocities in units of km\,s$^{-1}$. The  contours for the velocities
     in the left and middle panels  are in 50\,km s$^{-1}$ steps. In
     the right panel the contours are the CO(2--1) intensity (0$^{\rm th}$-order
     moment map) in a linear scale.}
              \label{fig:BAROLOvelfield}%
    \end{figure*}

\section{CO molecular gas emission}\label{sec:coldgas}

\subsection{Circumnuclear region}
Figure~\ref{fig:ALMAlargeFoV} (left panel) shows the integrated CO(2--1)
intensity map (the 0$^{\rm th}$-order moment map) covering a FoV of $10\arcsec
\times 10\arcsec$ with a beam size of $0.214\arcsec
\times0.161\arcsec$ at PA$_{\rm beam} = 42.0\degr$.  As previously shown by \cite{Schinnerer2000} using
PdBI observations at lower angular resolution, the circumnuclear cold molecular gas
shows an asymmetric ring-like  morphology with a
projected diameter of approximately $4-5\arcsec$ ($\sim
350\,$pc). This circumnuclear ring is also coincident with a stellar ring with
low  velocity dispersion, which \cite{Barbosa2006} interpreted as a
ring with recent star formation activity. The circumnuclear ring is
connected with the host galaxy disk through a large scale molecular bar, which extends for
10--15\arcsec \, to the northwest and southeast of the galaxy nucleus,
with the former part being brighter and in the direction of the
companion galaxy \citep{Meixner1990, Schinnerer2000}.

Our ALMA CO(2--1) observations 
further resolve the circumnuclear ring into several clumps none of which  
coincide with the AGN location derived from the peak of the unresolved
1.3\,mm continuum (see Section~\ref{subsec:continuum_morphology} and
below). The comparison of the CO(2--1) map with the {\it HST} F658N/F814W map in the
central $10\arcsec
\times 10\arcsec$ almost suggests an anti-correspondence between the
ionized gas (bright regions in this map) and the  molecular gas, except for the bright
\ion{H}{ii} region located approximately 4\arcsec \, southwest of the AGN. This
is likely explained because most of the cold molecular gas is in the
disk of the galaxy (see Section~\ref{sec:barolo}) while the
ionized emission in the circumnuclear region to the northeast is tracing outflowing material 
in  the ionization cone \citep{Mundell1995, Barbosa2009,
  Fischer2013}. In fact, the regions of high extinction to the west
and southwest in the near side of the galaxy, seen as dark lanes in
the F658N/F814W map and in the color maps presented by 
\cite{Chapman2000}, are all well traced by the CO(2--1) emission. 
This coincidence between  molecular gas and dust lanes is expected
and has been observed in other Seyfert galaxies
\citep{GarciaBurillo2005, Combes2013, Combes2014, AlonsoHerrero2018, Izumi2018}.

\subsection{Nuclear region}

Figure~\ref{fig:ALMAsmallCOmaps} zooms in on to the nuclear  $1\arcsec
\times 1\arcsec$ region of NGC~3227. The  left panels display in color the CO(2--1) and
CO(3--2) integrated line emission. These maps show a complex morphology
with  several peaks of emission. The higher angular resolution
CO(3--2) map reveals that none of these peaks are associated with the AGN
position. 
The peaks of the CO(2--1) emission appear to be
distributed along PA$_{\rm gas}=-40\degr$ extending for about
$1.5\arcsec$ (see also Figure~\ref{fig:ALMAlargeFoV}). The morphology
of the extended CO(2--1) emission is
similar to that seen in the $850\,\mu$m map when degraded to the same
resolution (top panel of Figure~\ref{fig:ALMAspectralindex}). 
The CO(3--2) morphology in the nuclear region is similar to that of the hot molecular gas traced by the
VLT/SINFONI adaptive optics near-infrared H$_2$ 1--0 S(1) line at $2.122\,\mu$m
\citep{Davies2006}. We reproduce this map in Figure~\ref{fig:SINFONI}
(left). It has a FWHM angular resolution of 0.085\arcsec,  which is
similar to that of the ALMA CO(3--2) map.
This indicates that near the AGN position these
CO transitions are tracing warmer gas than in other environments
(e.g., star forming regions), as also found in NGC~1068
by \cite{Viti2014}. 

The  CO(3--2) intensity map  resolves the inner
0.5\arcsec \, ($\sim 37\,$pc) emission into two main regions  with the northwest
clump being brighter of the two. In
projection the AGN is located approximately in between these two
CO(3--2) nuclear peaks. We
note that even though the AGN does not coincide with any of the
CO(3--2) clumps, there is molecular gas emission at the
AGN location, as we shall see in Section~\ref{sec:molgascolden}.
This nuclear molecular gas emission is also elongated similarly to that of
the nuclear disk $850\,\mu$m and 1.3\,mm continuum
emission. Furthermore, the superposition of the 1.3\,mm emission
contours on the CO(3--2) maps  
(Figure~\ref{fig:ALMAsmallCOmaps}) reveals that the faint 1.3\,mm continuum
emission also traces the innermost  molecular gas clumps.
There is also CO(3--2)  emission in
the direction of the polar dust/radio jet to the northeast of the
AGN. However, to the southwest of the
AGN there appears to be a region where some of the gas might have been
evacuated. In this region there is evidence of the 
presence of shocks \citep{Schoenell2019}. Based on the high velocity dispersion of the hot molecular 
gas in this region \cite{Davies2006} interpreted the hot molecular gas
emission as due to a highly turbulent rotating disk \citep[see
also][]{Hicks2009}.

 \begin{table*}
\caption{Summary of the $^{\rm 3D}$BAROLO model parameters.}             
\label{tab:BAROLOmodels}      
\centering                          
\begin{tabular}{c c c c c}        
\hline\hline                 
Model Name &  Large disk& \multicolumn{3}{c}{Nuclear region} \\
 &      component & component & parameters & type\\
  \hline                        
Disk  & fixed$^*$ & none & ... & ...\\
Disk + nuclear {\it slight} warp  &  var$=\pm 10\degr$&  warp  &
 $i_{\rm nuc}=+30\degr$, PA$_{\rm nuc-MAJ}=320\degr$  & var$=\pm 10\degr$\\
Disk + nuclear warp & fixed$^*$ &  warp & $i_{\rm nuc}=-30\degr$,
                                          PA$_{\rm nuc-MAJ}=320\degr$ & fixed\\
Disk + nuclear outflow & fixed$^*$ & non-circular  & $V_{\rm RAD } =180$km
                                                s$^{-1}$ & fixed\\
Disk + nuclear warp + outflow & fixed$^*$ & warp + non-circular&
                                                                 $i_{\rm
                                                          nuc}=-30\degr$,
                                                          PA$_{\rm
                                                          nuc-MAJ}=320\degr$\\
 & & &                                                          $V_{\rm RAD
                                                          } =180$km
                                                          s$^{-1}$ & fixed\\
  \hline                                   
\end{tabular}

Note.--- The models are those in Figure~\ref{fig:ALMACO21pv}, from top
to bottom. $^*$When fixed, the large scale disk parameters were  PA$_{\rm
  MAJ} = 152\degr$ and $i_{\rm disk} = 52\degr$ (see text). When not
fixed, they were allowed to vary around these values. For the nuclear
region we list the component name, the adopted parameters and whether
they were allowed to vary or not.

 \end{table*}

\section{CO(2--1) and CO(3--2) 
  kinematics}\label{sec:barolo}


In this section we study the molecular gas kinematics of the nuclear and
circumnuclear regions  of NGC~3227 (central $10\arcsec \times 10\arcsec$) using our
ALMA CO(2--1) and CO(3--2)
observations. 

Figure~\ref{fig:BAROLOvelfield} (left panel) shows that
in the central $\sim 4\arcsec \times 4\arcsec$ (a few hundred parsecs)
the molecular gas follows circular motions. 
The CO(2--1) velocity field in this figure is  noticeably similar to
that of the 1--0 S(1) H$_2$ line \citep{Davies2014}.  
However, at radial distances greater than 2-3\arcsec,  the CO(2--1) and H$_2$ velocity fields 
also show clear deviations from circular motions (see Section~\ref{subsec:inflow}). The stellar
kinematics on the other hand shows
relatively well ordered motions on these scales \citep{Barbosa2006,
  Davies2014}, although with high velocity dispersion
\citep{Davies2006} in the nuclear region. Within the inner 1\arcsec \,
the CO(2--1) and CO(3--2) transitions show very complex kinematics (see the mean velocity fields in the right panels of
Figure~\ref{fig:ALMAsmallCOmaps}). The motions do not appear to follow
the large scale velocity field except at the edges of the FoV. To
the southeast (approximately 0.2\arcsec) of the AGN position and along the direction of the nuclear disk,
there is a region with apparent
counterrotation. In the direction of the ionization cone axis there are also clear deviations from
circular motions. The hot molecular gas kinematics (see the VLT/SINFONI map in the right
panel of  Figure~\ref{fig:SINFONI}) traced with the 
1--0 S(1) H$_2$ line \citep{Davies2006} shows again a remarkable resemblance
with the CO(2--1) and CO(3--2) kinematics.  These authors also noted the high velocity dispersions and interpreted the
observations as due to the presence of a nuclear thick disk.

\begin{figure*}
   \centering
   \vspace{-0.4cm}

  \includegraphics[height=6.75cm]{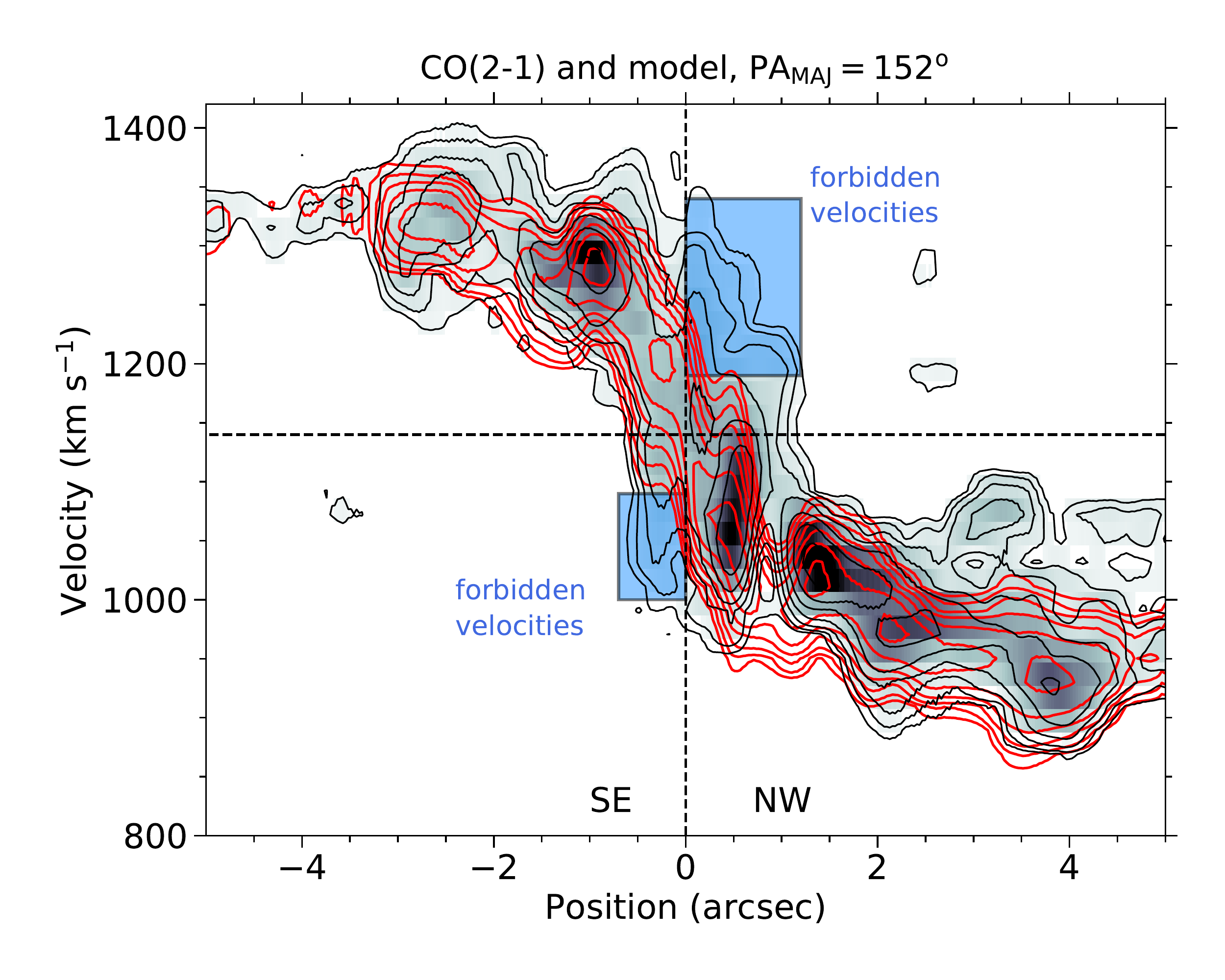}
  \includegraphics[height=6.75cm]{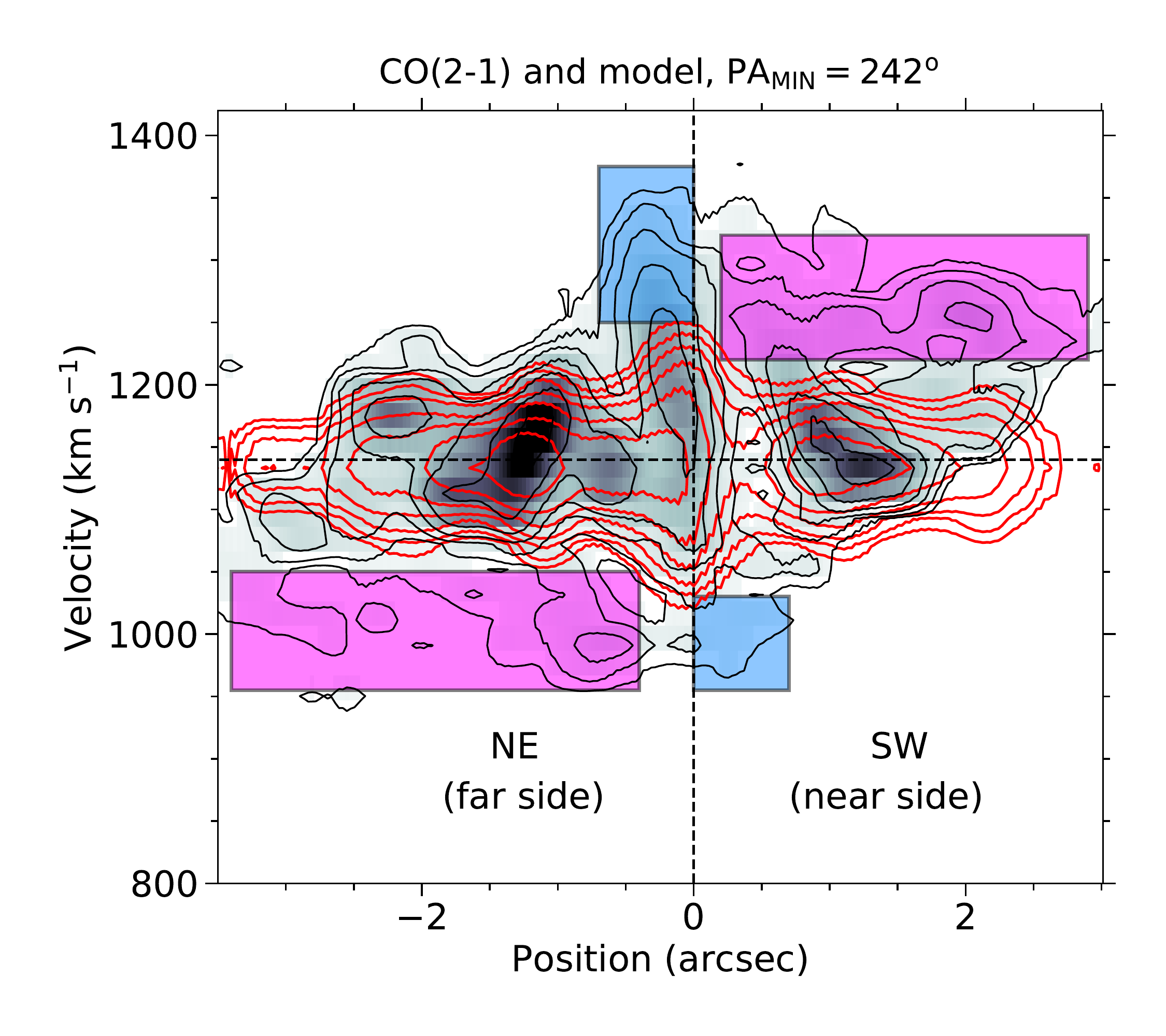}

\vspace{-0.2cm}
  \includegraphics[height=6.75cm]{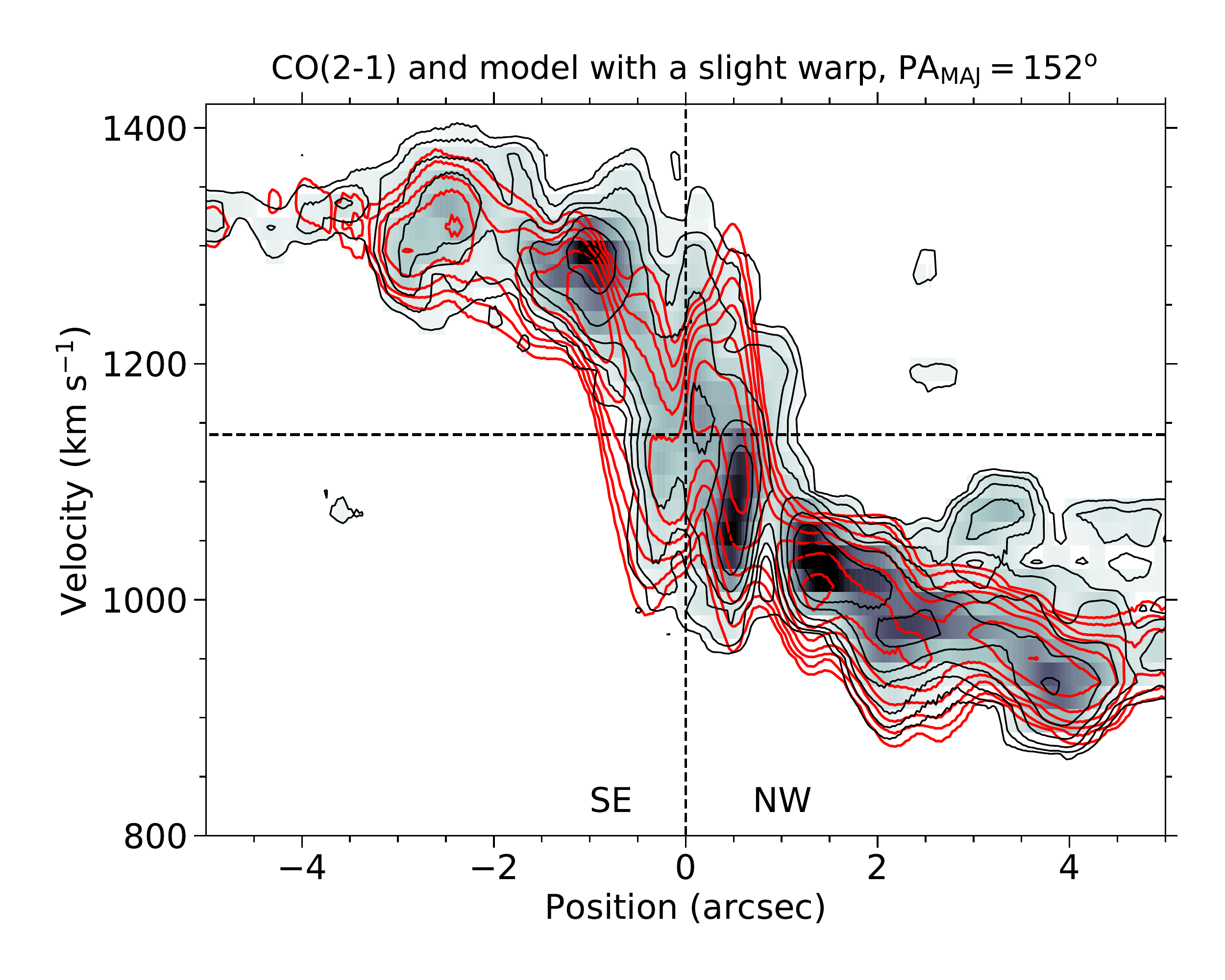}
  \includegraphics[height=6.75cm]{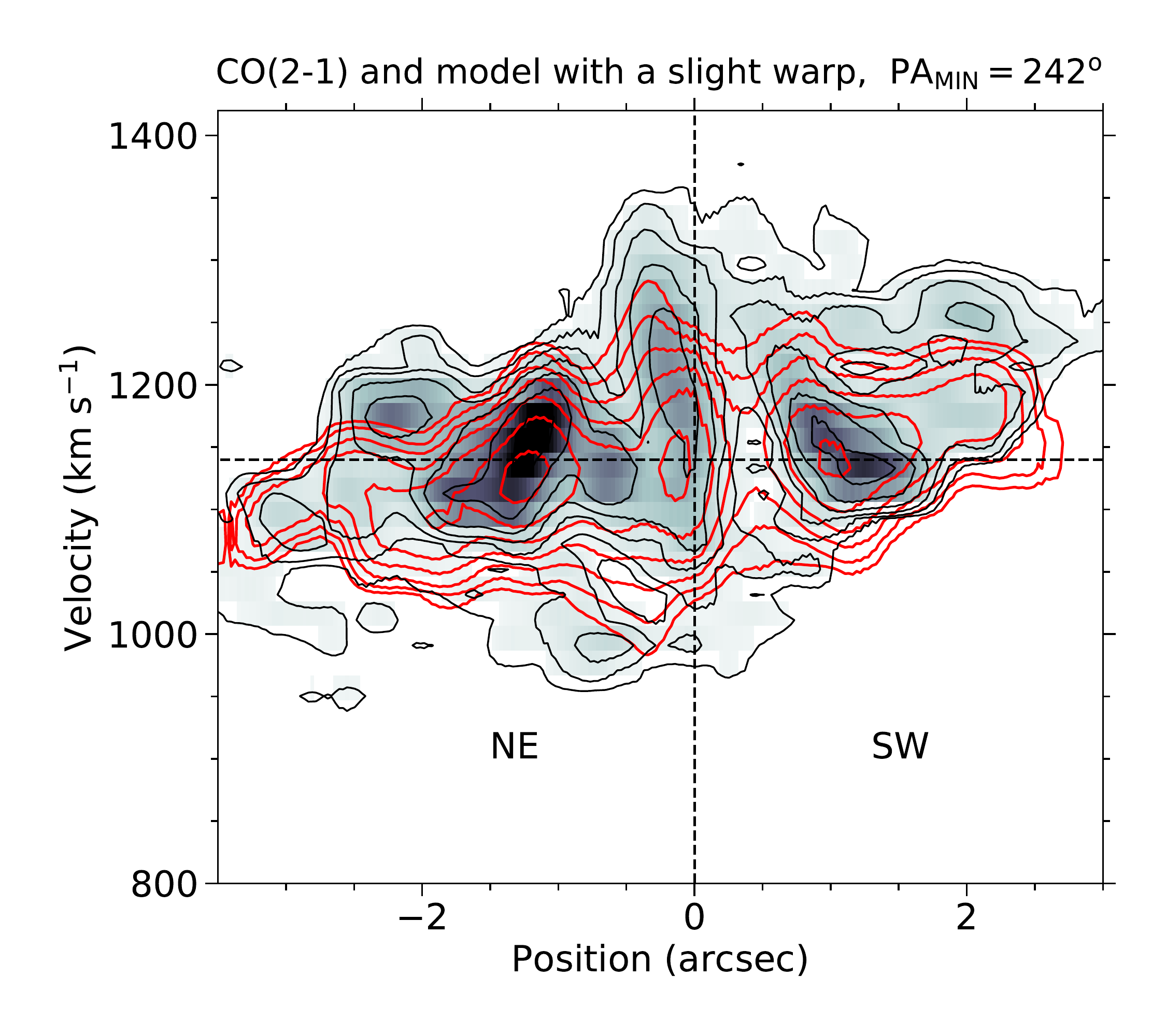}

\vspace{-0.2cm}
 \includegraphics[height=6.75cm]{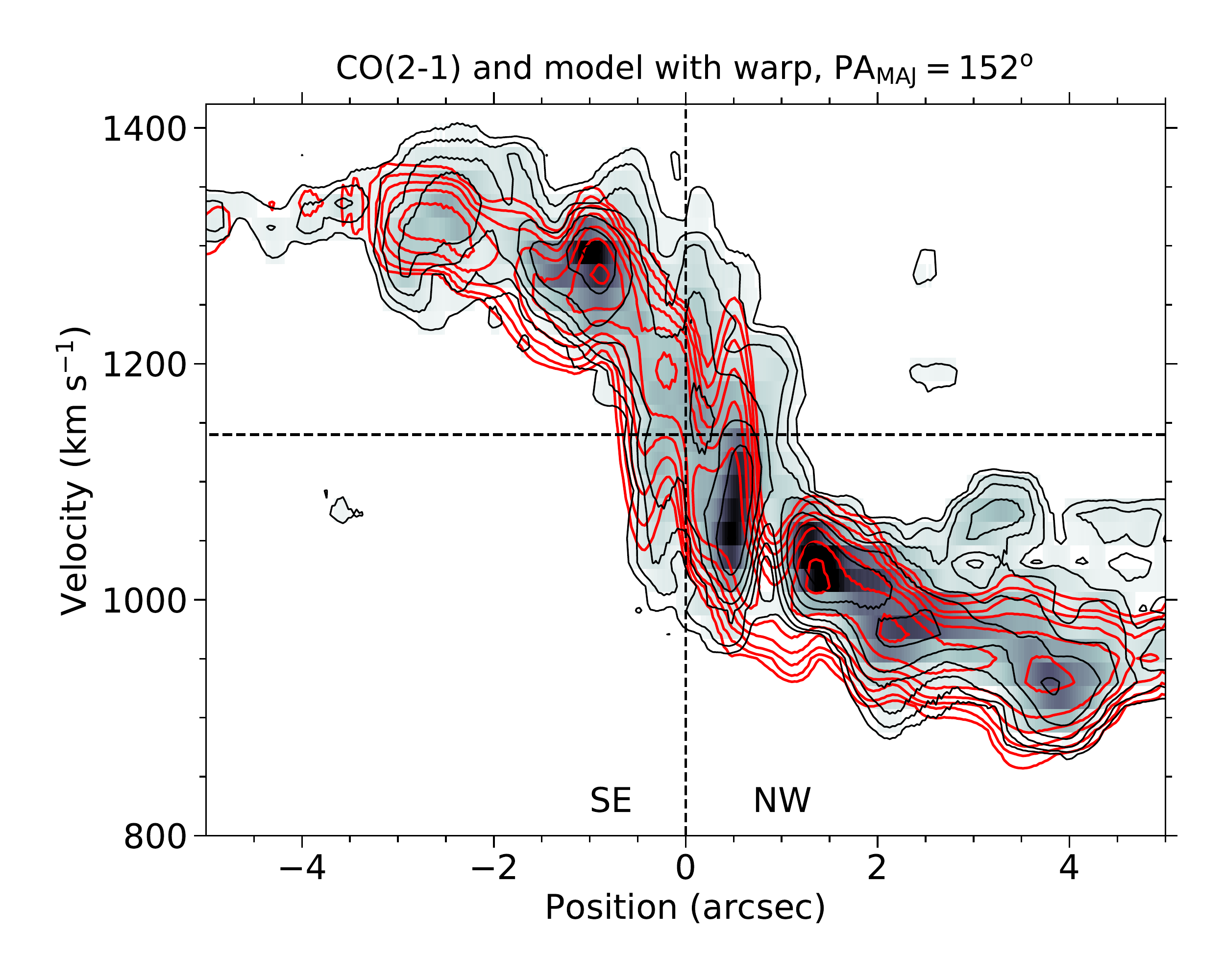}
 \includegraphics[height=6.75cm]{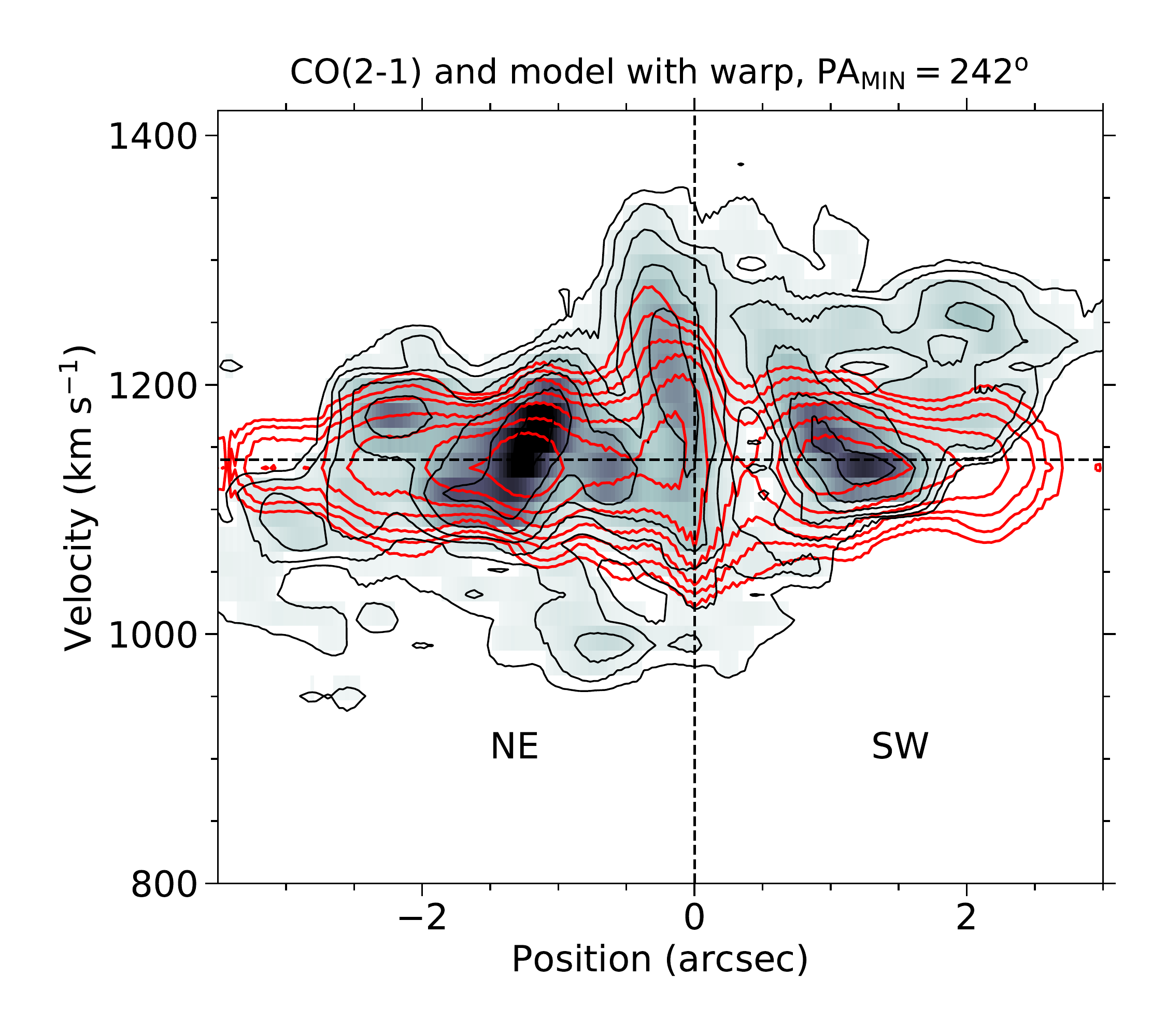}

  
  \vspace{-0.3cm}
  \caption{In grey-scale and black contours are the observed CO(2--1)
    p-v diagrams along the kinematic major
    axis (left panels) and minor  axis (right panels) of the
     host galaxy disk. We only
     show fluxes above 5$\sigma$. In all the panels the red
     contours represent the 
$^{\rm 3D}$BAROLO 
     models as follows (see also Table~\ref{tab:BAROLOmodels}). The first row is a rotating
     disk, the second row is a
     rotating disk  with a {\it slight} nuclear warp, the third row is a rotating disk + a warp model consistent
     with the orientation of the ionization cone, the fourth
     row is a rotating disk with a nuclear radial component, and the
     fifth row is a rotating disk with a nuclear warp + radial component. The model
     contour levels are identical to the contour level of the observations. The vertical dotted black line marks the AGN
     position and the horizontal dotted black line the derived systemic
     velocity. In the top panels, we also highlight nuclear regions of {\it forbidden} velocities
     for a purely rotating disk. Along the minor axis the blue boxes
     indicate the nuclear outflow and the magenta ones 
       regions with streaming motions due to the large scale stellar bar.}
              \label{fig:ALMACO21pv}%
    \end{figure*}

\begin{figure*}
\setcounter{figure}{7}
   \centering
   \vspace{-0.4cm}

  \includegraphics[height=6.75cm]{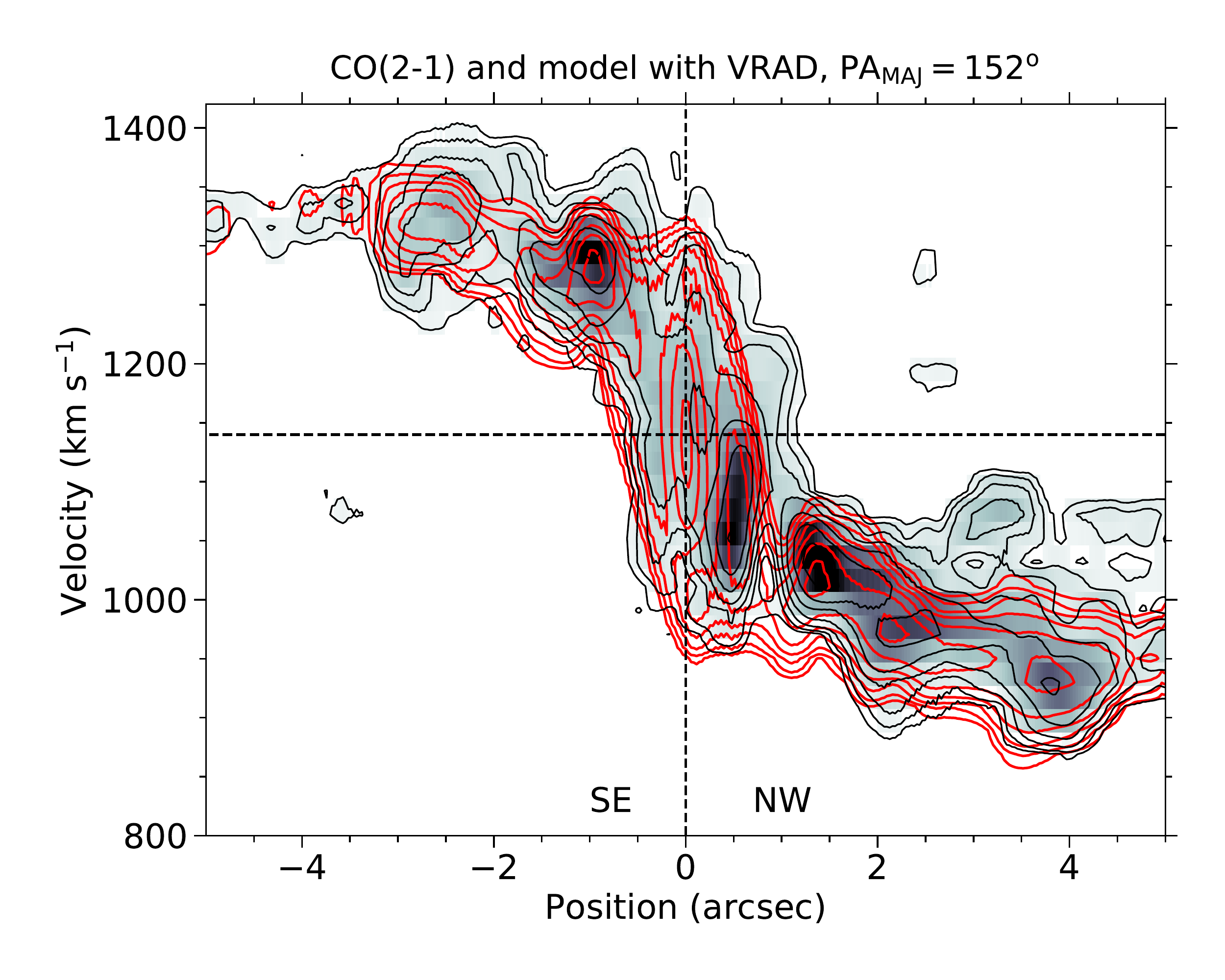}
  \includegraphics[height=6.75cm]{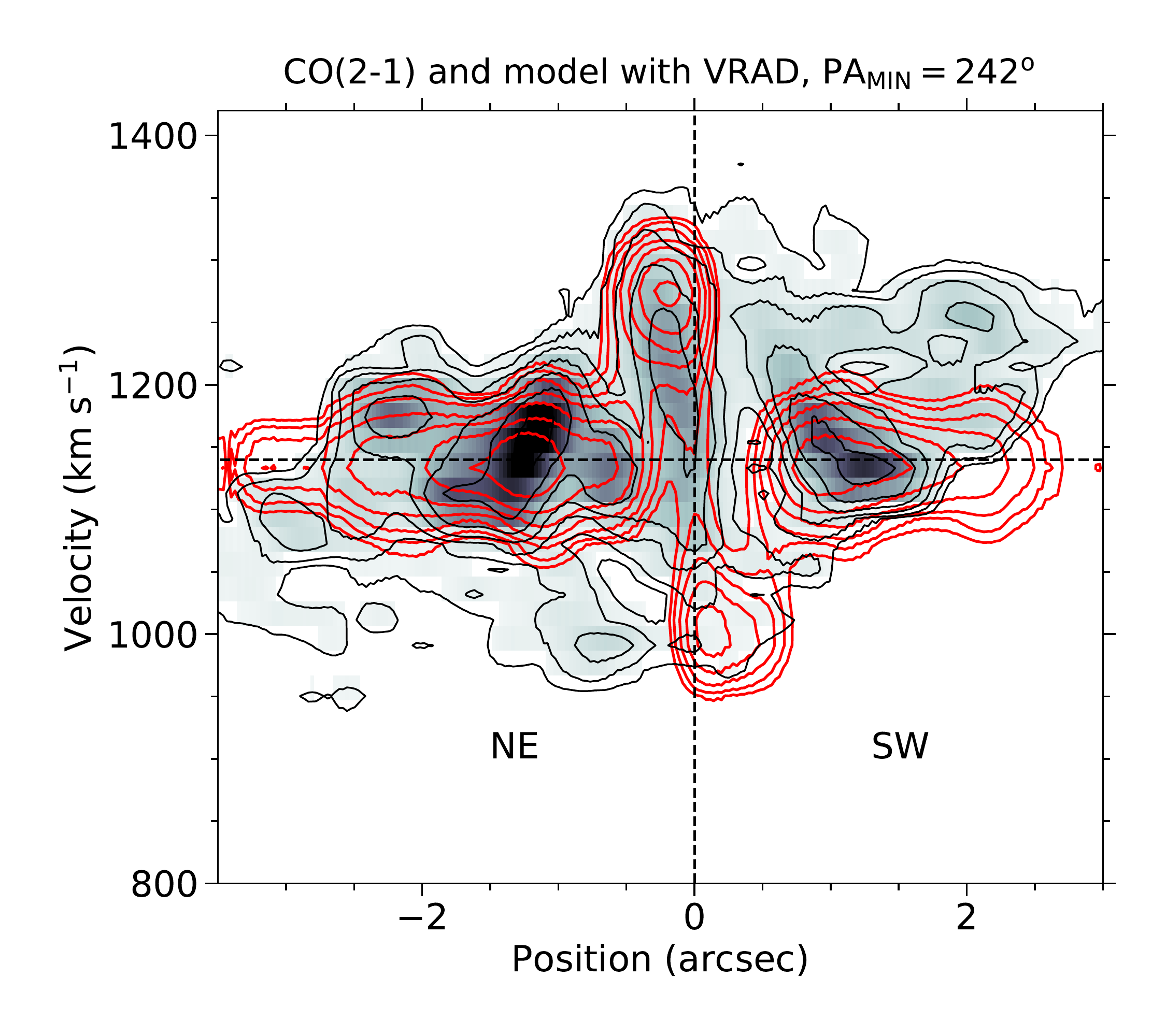}

\vspace{-0.2cm}
  \includegraphics[height=6.75cm]{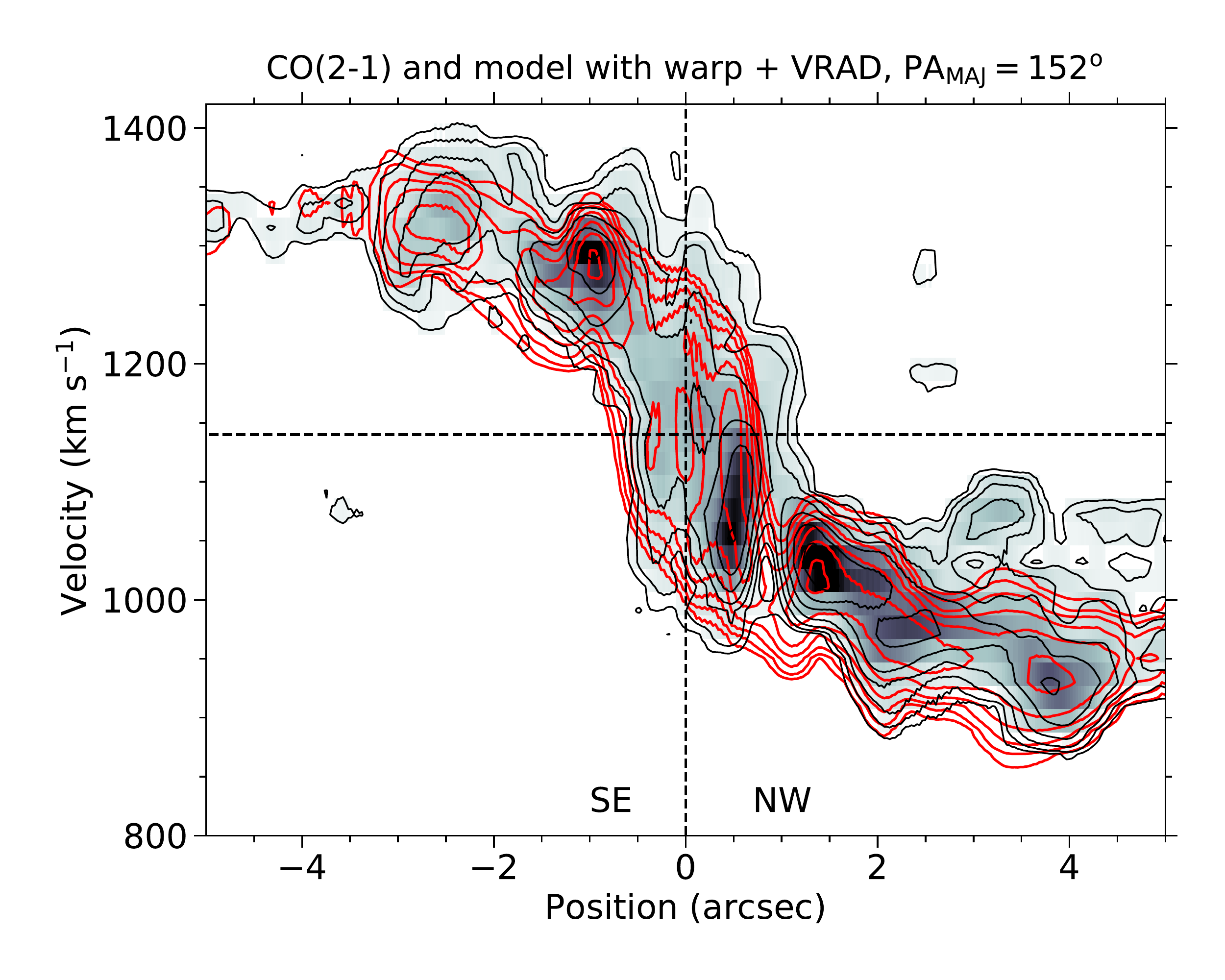}
  \includegraphics[height=6.75cm]{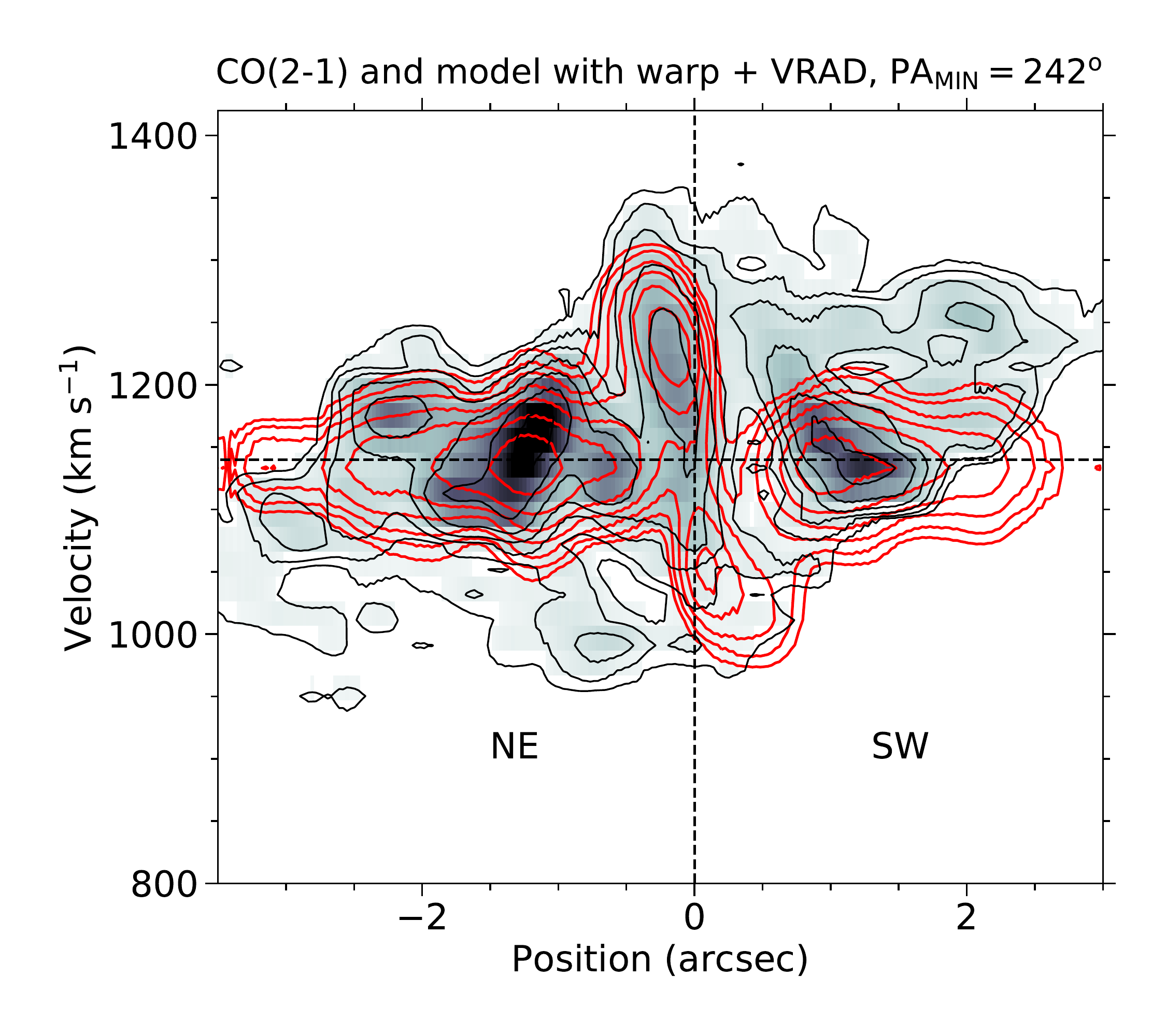}

  \vspace{-0.3cm}
  \caption{Continued.}
    \end{figure*}

\begin{figure*}
   \centering
  \includegraphics[height=7.5cm]{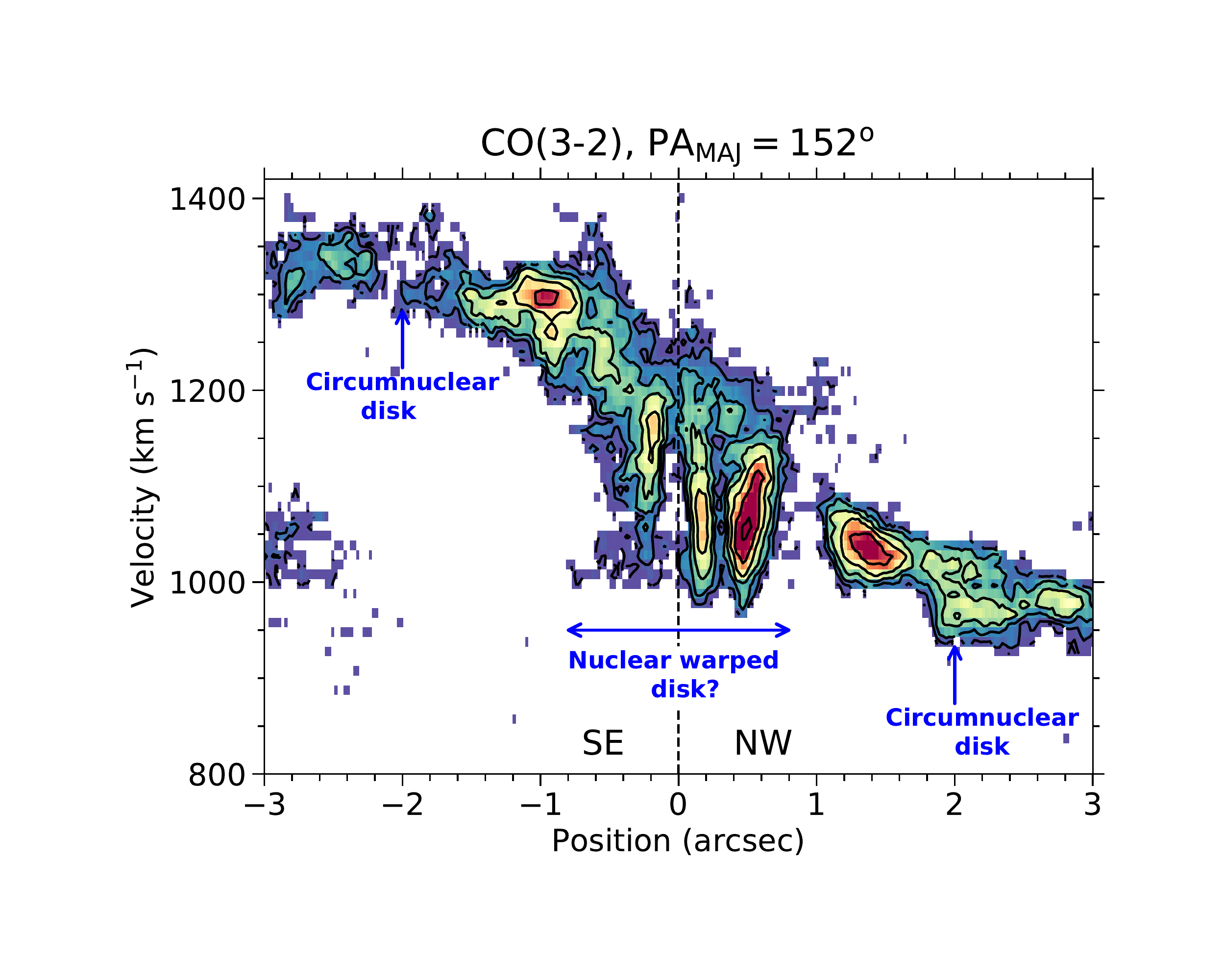}
  \includegraphics[height=7.5cm]{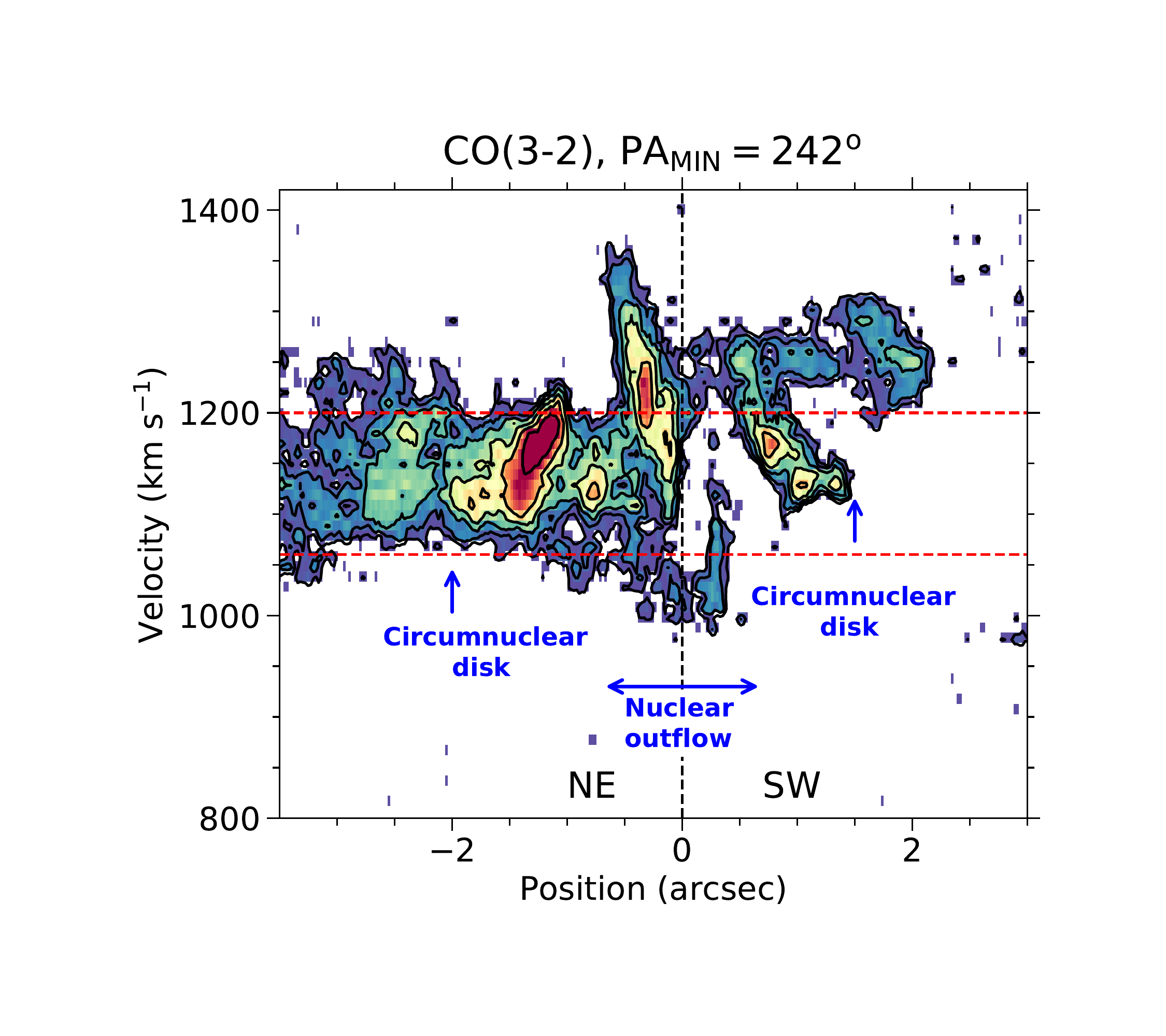}

  \vspace{-1cm}
  \caption{In color are the observed CO(3--2) p-v
    diagrams along the kinematic major
      axis (left) and minor  axis (right) of the
     host galaxy disk.  We only
     show fluxes above 3$\sigma$. The black
     contours are fluxes in a linear scale. The black dotted vertical line
     marks the AGN position whereas the horizontal red lines mark the
     approximate velocity virial range of the simple rotating disk model
      (see top panel of Figure~\ref{fig:ALMACO21pv}).}
              \label{fig:ALMACO32pv}%
    \end{figure*}

To understand the kinematics and the strong deviations from 
circular motions, we model the CO(2--1) data cube with
$^{\rm 3D}$BAROLO \citep{DiTeodoro2015}.  This code was
developed to fit 3D tilted-ring models to emission-line data cubes using simple rotating
disks. We first started by fitting a simple $^{\rm 3D}$BAROLO  model of a rotating disk
allowing the following parameters to vary: the systemic velocity
($v_{\rm sys}$), the disk inclination ($i_{\rm disk}$) and
PA\footnote{$^{\rm 3D}$BAROLO defines the position
angle of the major axis on the receding half of the
galaxy, taken anticlockwise from the north direction on the sky and
the inclination with respect to the observer ($90\degr$ for an edge-on
view).} of the kinematic major axis (PA$_{\rm MAJ}$), the circular
velocity,  and the velocity dispersion ($\sigma_{\rm gas}$). For the kinematic center we
used the position of the AGN as that of the unresolved component from
the two component fit to the 1.3\,mm
continuum emission (see Section~\ref{subsec:continuum_morphology}). 
We first fitted the
central few arcseconds where clear circular motions are observed and 
 derived an average value of  $v_{\rm sys} =1140$\,km s$^{-1}$. With
 the systematic velocity fixed, we fitted the central $10\arcsec
 \times 10\arcsec$ and derived average values of the major axis PA and inclination of
the host galaxy disk
of PA$_{\rm MAJ}=152\degr$  and $i_{\rm
  disk}=52\degr$, respectively.  These are in good 
agreement with those inferred from the stellar kinematics over similar
physical scales \citep{Barbosa2006, Davies2006, Davies2014, Hicks2009}. 
The next step was to fix the values of the galaxy disk PA and inclination and the
systemic velocity and fit only the velocity dispersion and circular velocity. The resulting
$^{\rm 3D}$BAROLO model together with the CO(2--1) mean velocity field 
(1$^{\rm st}$-order moment map) are presented in
Figure~\ref{fig:BAROLOvelfield}. 

In the following subsections we will
use this simple rotating disk model as our base model to first look for departures from
circular motions and then to add nuclear components to try to
reproduce the observed kinematics.

\begin{figure*}
   \centering
  \includegraphics[height=6.75cm]{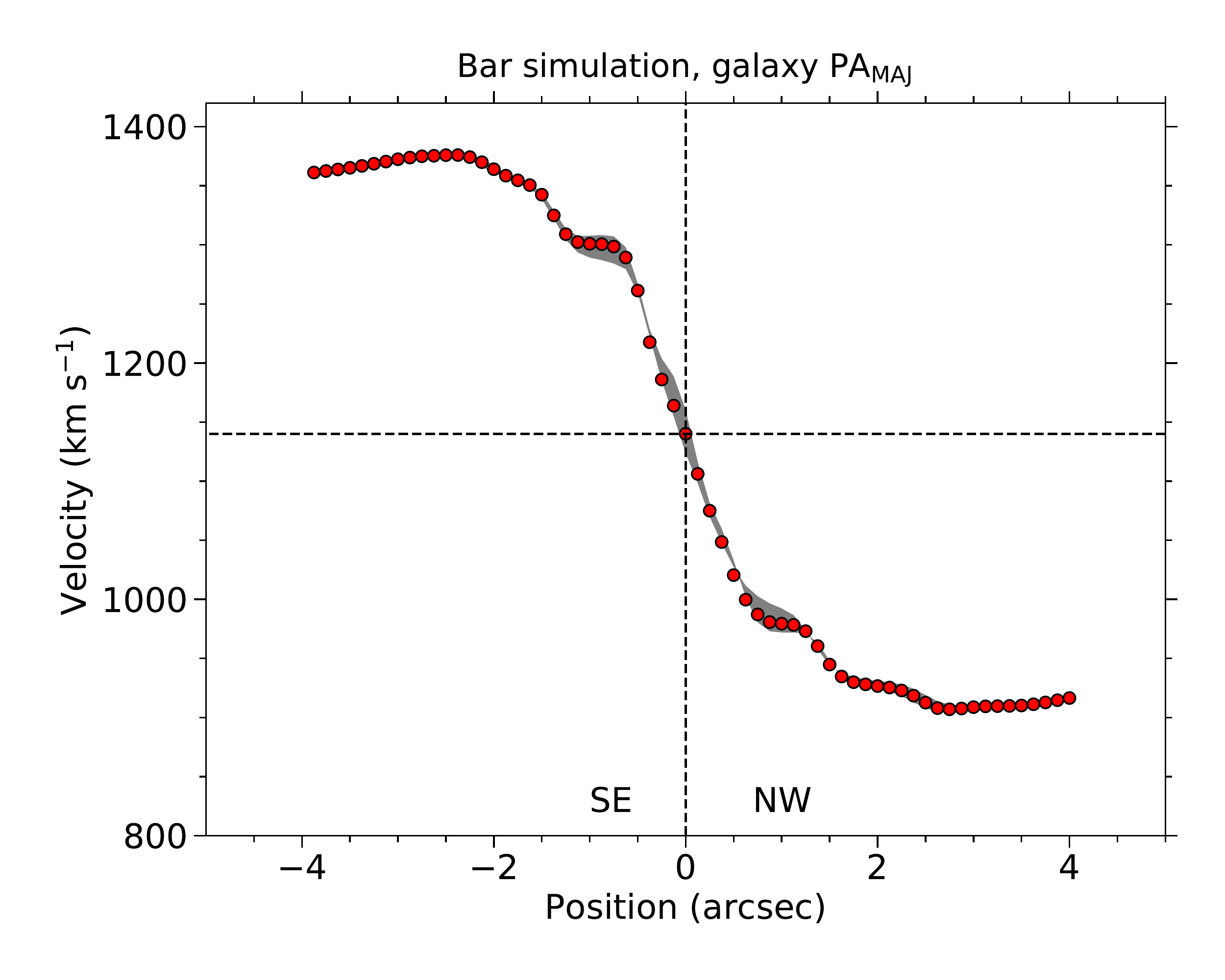}
  \includegraphics[height=6.75cm]{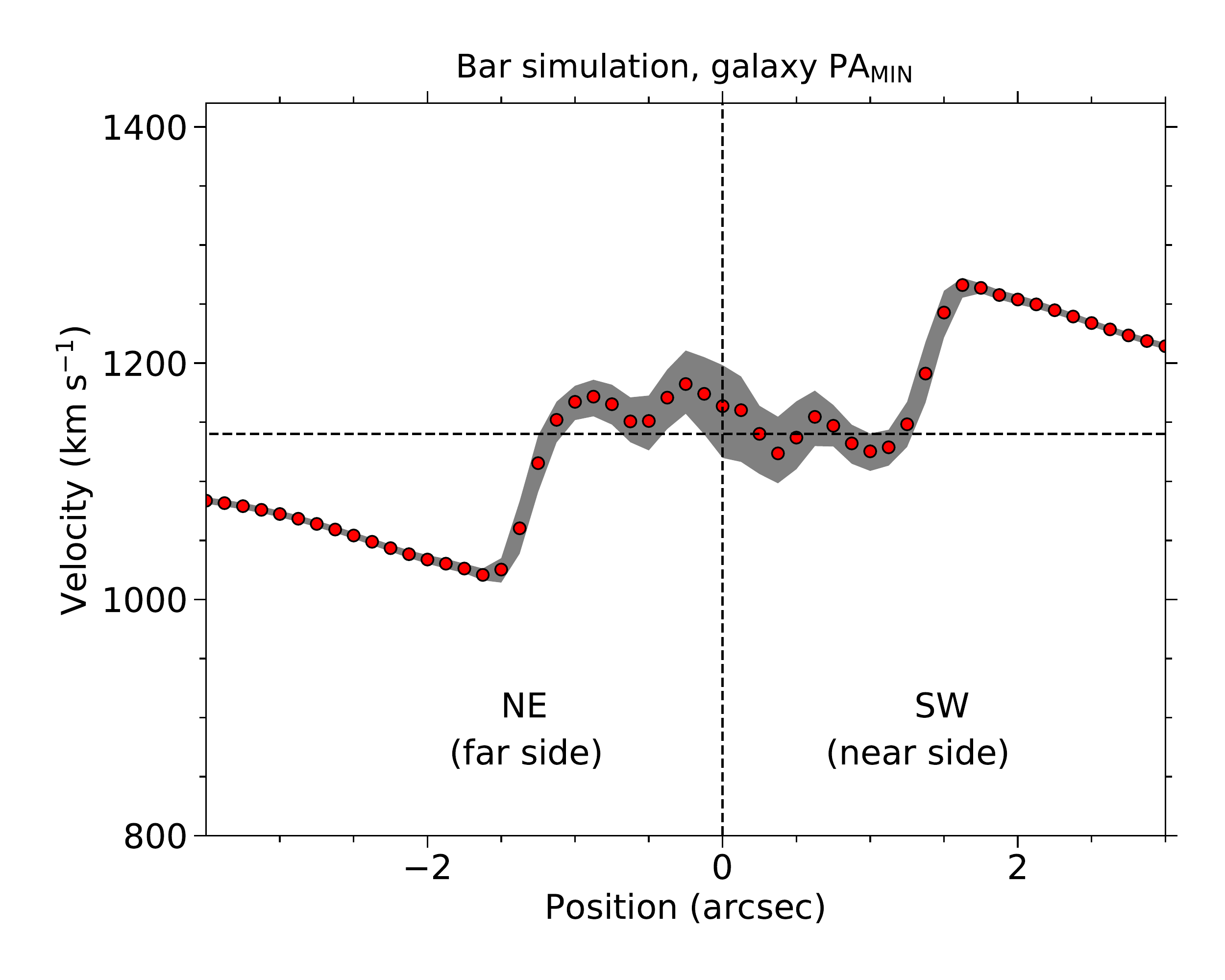}
\vspace{-0.3cm}
  \caption{ Gas line-of-sight velocity versus distance from the center along the galaxy disk kinematic major (left panel) and minor
    (right panel) axes. The gas line-of-sight velocity field
    corresponds to the 0S20r  simulation 
of a strong bar \citep{Maciejewski2004}, and adapted to our ALMA
resolution as well as to the NGC~3227 disk geometry, and observed CO(2--1) velocities.
In each panel the cut passing through the center is plotted with the
dot symbols. We also extracted two additional cuts  on both
sides of the center at $\pm 0.1\arcsec$ which are represented with the
shaded area. These plots can be directly compared with the observed CO(2--1)
p-v diagrams in the top panels of Figure~\ref{fig:ALMACO21pv}. }
              \label{fig:barmodel}
    \end{figure*}

\subsection{Streaming motions}\label{subsec:inflow}

We constructed  a  CO(2--1)  residual mean velocity field map (right panel of
Figure~\ref{fig:BAROLOvelfield})  
by subtracting the $^{\rm 3D}$BAROLO rotating disk model from the observed mean velocity field. 
On circumnuclear scales, the model 
reproduces the  mean velocity field except along
the minor kinematic axis. There are 
strong velocity deviations from
circular motions  at approximately projected $r>2-3\arcsec$  
along this axis, as well as in the nuclear region (see below). The velocity residuals, which are of the order
of $100\,{\rm km\,s}^{-1}$ or more, 
are positive to the northwest and west (projected radial distances greater than
$\sim 200\,$pc)  and negative to the 
southeast (projected radial distances greater than $\sim 300\,$pc). 
\cite{Davies2014}
observed similar velocity residuals on scales of a few arcseconds from
the AGN when modeling the kinematics of
the near-infrared H$_2$ line at $2.12\,\mu$m with a rotating disk. 


We can also compare the observed CO(2--1)
position-velocity (p-v) diagrams along the major and minor kinematic axes of
the host galaxy disk (shown in Figure~\ref{fig:ALMACO21pv} in grey
scale and black contours) with the fitted $^{\rm 3D}$BAROLO model
of a rotating disk (red contours, top panel). Along the
kinematic major axis the $^{\rm 3D}$BAROLO model with the rotating
disk fits
fairly well the rotation curve of the host galaxy disk for projected radial
distances $\sim 1-5\arcsec$. However it fails to 
reproduce it at radial distances from the nucleus of
less than approximately 1\arcsec, where we find emission  from {\it forbidden}
velocities for a simple rotating disk.  
Along the kinematic minor axis of the host galaxy disk,
again the $^{\rm 3D}$BAROLO rotating disk model
reproduces only the observations outside the nuclear regions. However, 
at radial distances greater than 1\arcsec, we can clearly see the 
  non-circular motions
 to the southwest and northeast parts of the galaxy, where we identified the strong
mean-velocity residuals in Figure~\ref{fig:BAROLOvelfield}. Along this position angle
the inflow appears to be stronger to the
southwest of the AGN. 

Figure~\ref{fig:ALMACO32pv} shows the same observed p-v
diagrams along the kinematic major and minor axes of the host galaxy
disk for the CO(3--2) transition at higher angular resolution (see
Table~\ref{tab:ALMAmoleculargas}). Qualitatively we observe the same evidence
for  non-circular motions along the kinematic minor axis of the host
galaxy especially to the southwest of the AGN and as close as
0.5-0.7\arcsec \, (projected distances 35-50\,pc) from the AGN
location (Figure~\ref{fig:ALMACO32pv}, right
panel). We can also see the clear
kinematic decoupling of the nuclear emission in the p-v
diagram along the major kinematic axis taking place at projected
radial distances of less than $\sim 0.7\arcsec$.

  As discussed in the Introduction, NGC~3227 presents a large
  scale stellar bar \citep{Mulchaey1997} with a measured position
angle of  PA$_{\rm  bar}=150\degr$ and a length greater than 80\arcsec. 
Based on the bar properties, \cite{Davies2014} used a hydrodynamical simulation  developed by
\cite{Maciejewski2004} to illustrate the effects of the presence of a
strong stellar bar in the  gas kinematics of NGC~3227.  They showed that the simulated
line-of-sight velocity field reproduced well both the
circular motions and the
presence of streaming motions at the outer part of  the 
ring, as observed in their H$_2$  map and also our CO(2--1) map. The streaming motions are expected in and
near the
dust lanes entering the circumnuclear ring, which is exactly what we
observe in our maps
(see Figures~\ref{fig:ALMAlargeFoV} and
\ref{fig:BAROLOvelfield}).  

Here we evaluate if  the perturbation from a large scale bar alone is
responsible for the  CO(2--1) non-circular motions observed on
different physical scales in NGC~3227. To do so, we 
used the same \cite{Maciejewski2004}  bar simulation (model termed
0S20r) used by
\cite{Davies2014} and adapted it to the disk geometry fitted with $^{\rm
  3D}$BAROLO. Then we smoothed it to the angular
resolution of our ALMA CO(2--1) data (assuming a circular Gaussian
with a FWHM of 0.2\arcsec) 
and scaled it to the observed velocities and derived $v_{\rm sys}$. To compare with the observed CO(2--1) p-v
diagrams,  we extracted plots of the line-of-sight velocity against
the distance from the center along the major and minor axes from the simulated
velocity field of the gas. The comparison of Figure~\ref{fig:barmodel} and
Figure~\ref{fig:ALMACO21pv} (top panels) shows that the bar simulation
model is able to reproduce the line-of-sight velocity along the major
 axis as well as the
streaming motions in regions along the minor axis at $\sim 1.5-3.5$\arcsec\, to the northeast and
southwest of the AGN  (marked with the magenta rectangles in
Figure~\ref{fig:ALMACO21pv}). Since at these radial
distances we are inside the large scale bar corotation, the streaming motions
are associated with material inflowing near the leading edges of the
bar. The bar model cannot, however  reproduce the amplitude of 
 the forbidden velocities
along the major axis (marked with the blue rectangles in
the left top panel of Figure~\ref{fig:ALMACO21pv}).

Along the minor axis, the bar model cannot reproduce the
strong non-circular motions in the nuclear region ($r<1\arcsec$). 
This finding is typical for the gas flow in barred potentials, which
becomes close to circular inwards 
from the nuclear ring \citep[e.g., ][]{Piner1995}. In model 0S20r, in
addition to the inflow along the bar, the bar triggers nuclear spiral
shocks, which can be revealed in the line-of-sight velocity residuals
\citep[e.g., ][]{StorchiBergmann2007, Davies2009}, and are
manifested by the "wiggles" at radial distances $r<1$\arcsec\, in the nuclear
region along the model minor axis p-v diagram (right panel of
Figure~\ref{fig:barmodel}). However, the amplitude of these residuals is not sufficient
to explain the observed velocities along the minor axis of NGC~3227.


\subsection{A nuclear warp?}\label{subsec:nuclearwarp}

\cite{Schinnerer2000} observed similar non-circular motions in
  the nuclear region of NGC~3227 using
lower resolution CO(2--1) observations. They first attempted to model them
with a nuclear bar but
could not reproduce their observed p-v diagrams in the inner $\sim
1$\arcsec. They explored a second possibility using a circumnuclear disk that is warped 
at approximately radial distances of 0.7-1\arcsec \, from the nucleus and it  is orthogonal
to the host galaxy plane in the innermost region. They found that this
model reproduced better the CO(2--1) observations and the geometry
would agree  with that required from the kinematics of the
ionized gas of NGC~3227. Indeed, \cite{Fischer2013} derived 
an angle between the 
axis of the ionization cone and the normal to the host galaxy disk
of $\beta=76\degr$ \citep{Fischer2013}. Therefore, this would also imply 
a {\it drastic} gradient of the axis of the rotating molecular gas as
it approaches the nuclear collimating structure.

In this subsection we explore briefly  with $^{\rm 3D}$BAROLO and our
high angular resolution ALMA data the possibility of the presence of
a nuclear warp. To do so,  we took the initial model fitted with
$^{\rm 3D}$BAROLO and refitted the nuclear region (radial distances of
less than 1\arcsec) to allow for the
presence of a {\it slight}  geometric warp, in the sense of a small
tilt between the inclination of the host galaxy and the nuclear region.
This is because we do not know the scales on which this change of tilt
happens, that is, whether on the scales probed by our ALMA CO(2--1) data or much
further in. As before, we fixed the kinematic
center and $v_{\rm sys}$. For the nuclear warp we started with
initial values of $i_{\rm nuc} = +30\degr$ and PA$_{\rm MAJ-nuc} =
320\degr$. The change of almost 180\degr \, in the PA with respect to
that of the host galaxy disk (that is, the twist) is necessary to fit the 
counterrotation seen in the nuclear region and to agree with the orientation of
the nuclear CO(2--1) clumps (see Figure~\ref{fig:ALMAsmallCOmaps}). Using
these initial values we allowed 
$^{\rm 3D}$BAROLO to vary the nuclear and disk PA and inclination by $\pm 10\degr$. As
can be seen from the second row of Figure~\ref{fig:ALMACO21pv}, in the
p-v diagram along the kinematic major axis the presence
of a nuclear warp starts populating some of the {\it forbidden}
velocities. 

We run another $^{\rm 3D}$BAROLO model allowing for 
a more dramatic warp
to make it consistent with the inclination of the collimating
structure ($i_{\rm nuc} = -30\degr$) for the ionization cone. In other
words,  we assume that
the warp required takes place on the physical scales probed by our ALMA 
CO(2-1) data set. In this case we did not allow
the nuclear PA and inclination to vary but  kept them fixed (see
Table~\ref{tab:BAROLOmodels} for a summary).  The  model
p-v diagrams are rather similar (third row of
Figure~\ref{fig:ALMACO21pv}) to those of the {\it 
  slight} warp.  This suggests that with this data set we cannot determine the physical scales where this 
warp would be taking place.  We note that the model with the {\it slight} warp appears to
produce a marginally better fit to the nuclear region along the major
axis. This is because  the fitted velocity dispersion in the nuclear
region for the {\it slight} warp goes up to values 
$\sigma_{\rm gas} \simeq 50-100\,{\rm km\,s}^{-1}$ (but see
Section~\ref{subsec:nuclearoutflow}) while the drastic warp model only
fits values up to $\sigma_{\rm gas} \simeq 75\,{\rm km\,s}^{-1}$.
The effects of the increased velocity dispersion in the models with
the warp can also be seen in the
p-v diagrams along the  
minor axis of the model  when compared to the simple
rotating disk. Even so, the nuclear warps cannot reproduce the strong non-circular
motions to the northeast and southwest along the minor axis (which are
also seen in the hot H$_2$ gas velocity map in Figure~\ref{fig:SINFONI}) in the nuclear
region.

\subsection{Nuclear non-circular motions}
\begin{figure*}
   \centering

\hspace{-1cm}
  \includegraphics[width=8cm]{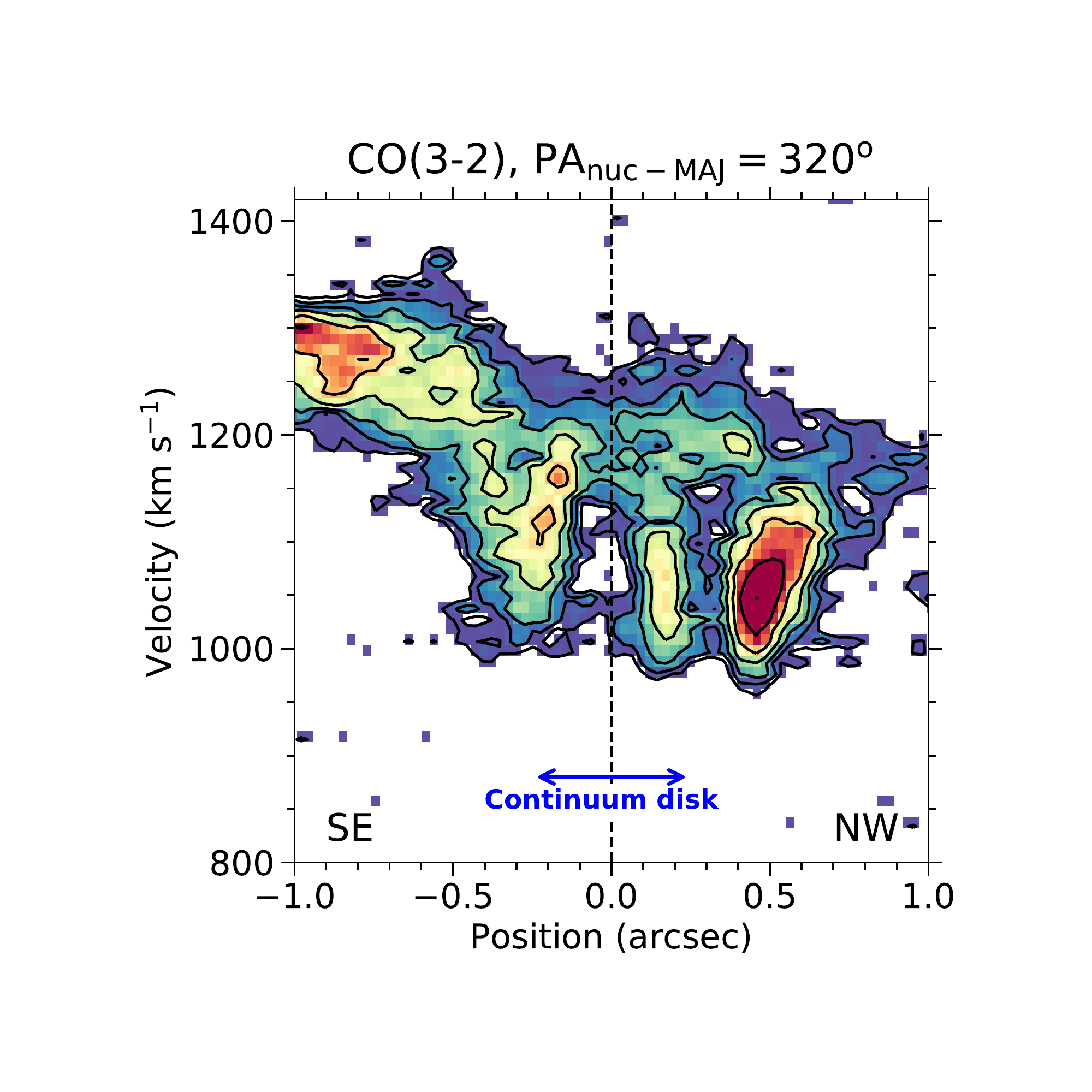}
  \includegraphics[width=8cm]{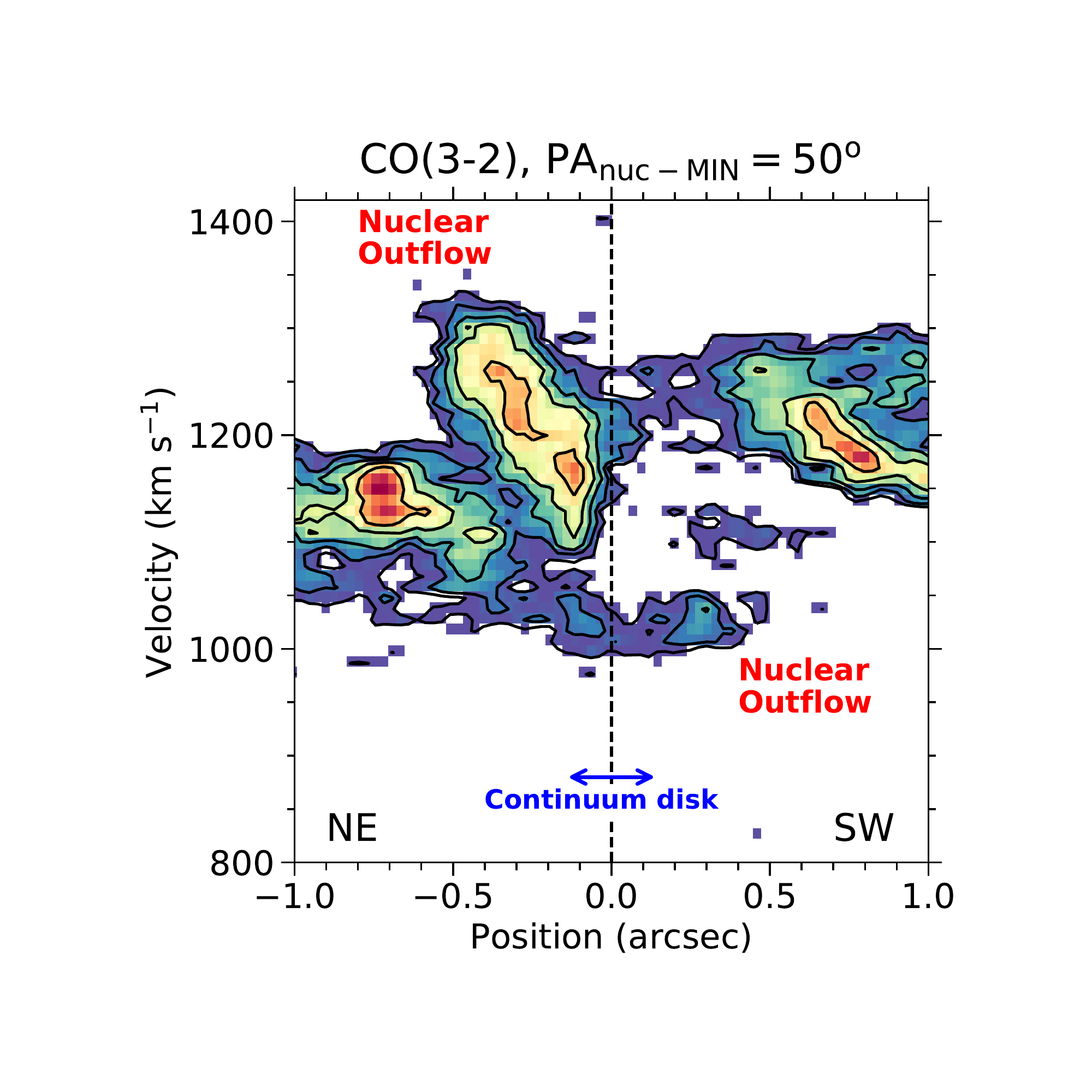}
  \vspace{-1cm}

  \caption{In color are the CO(3--2) p-v diagrams along the major
and      minor kinematic axes  of 
     the nuclear disk.  We only
     show fluxes above 3$\sigma$. The black
     contours are fluxes in a linear scale. The arrows show the approximate projected size
  of the $850\,\mu$m nuclear continuum emission.}
              \label{fig:ALMACO32pv_nucleardisk}%
    \end{figure*}

 The observed CO(3--2) p-v diagram along the minor axis of the nuclear disk
(PA$_{\rm nuc-MIN}=50\degr$) in
Figure~\ref{fig:ALMACO32pv_nucleardisk} clearly shows non-circular motions.
 The deviations from circular motions
reach line-of-sight velocities of
$(v-v_{\rm sys}) \sim +200\,{\rm km\,s}^{-1}$ (i.e., redshifted) to the
northeast of 
the AGN out to a projected distance $r=0.5\arcsec$, 
and  $(v-v_{\rm sys})\sim -150\,{\rm km\,s}^{-1}$ (i.e., blueshifted) to the southwest of
the AGN at a similar distance. Therefore, these radial motions are
observed to extend for approximately $1\arcsec$ ($\sim 70\,$pc) 
along the minor axis of the nuclear disk. These excess velocities with
respect to a purely rotating nuclear disk
are in agreement with the nuclear velocity residuals observed in hot
molecular gas using the near-infrared H$_2$ line  by
\cite{Davies2014}.  

To illustrate the effects of  strong non-circular motions in the nuclear
regions, as seen along the kinematic minor axis, we run another $^{\rm 3D}$BAROLO
model again using the initial rotating disk model and including a
radial velocity component in the nuclear region. Based on the
deviations discussed above we fixed radial velocity component in the
nuclear region ($r<1\arcsec$) to  $V_{\rm RAD}=180\,{\rm km\,s}^{-1}$ (see
Table~\ref{tab:BAROLOmodels}). We allowed  the rotational velocity and
the velocity dispersion to vary, and left the other parameters
fixed. The fitted model is shown in the
fourth row of Figure~\ref{fig:ALMACO21pv}. We can see that this 
model reproduces fairly well the p-v diagram along the
minor axis and in the nuclear region, as can be seen more clearly when
looking also at the right panel of
Figure~\ref{fig:ALMACO32pv_nucleardisk}. We note that this is not {\it
  the best fit} to the data, as there are also inflowing motions near the nuclear region seen to the southwest and northeast of the
AGN  (see above). However, 
 with $^{\rm
  3D}$BAROLO we cannot include simultaneously at a particular radial 
distance  radial velocity components with different signs. Thus we
include the dominant radial component in the nuclear region, which produces a better fit to the
nuclear part of the p-v diagram while keeping the fitted
velocity dispersion relatively low (approximately $\sigma_{\rm gas} = 20-30\,{\rm km\,s}^{-1}$).

As a final $^{\rm 3D}$BAROLO model, we combine a nuclear warp with a
radial component, with the parameters listed in Table~\ref{tab:BAROLOmodels}. Visually this last
 model (Figure~\ref{fig:ALMACO21pv}, 5th row) produces a better fit to 
 p-v diagrams along both the major and minor axes than just having the
nuclear warp or the radial component alone, respectively.
Additionally it does not need large nuclear gas velocity
 dispersions at the transition region of the warp.

\subsection{A nuclear molecular outflow}\label{subsec:nuclearoutflow}

In the previous subsection we showed that  non-circular motions along the minor axis
of the galaxy are certainly detected within the central 1\arcsec, and more importantly
along the minor axis. 
 If we assume that these motions are in the
plane of the putative nuclear disk (at $i_{\rm nuc}=-30\degr$), then
these motions would be interpreted as 
inflow. However, the inferred radial velocities (up to $150-200\,{\rm
  km\,s}^{-1}/\sin i_{\rm nuc}$) would be unrealistically high for a nuclear
inflow. For instance, the hydrodynamical simulations run to fit the
observations of the low luminosity AGN NGC~1097 required radial flows
of the order of $20-50\,{\rm km\,s}^{-1}$ \citep{Davies2009}. 
  Even in the presence of a nuclear bar, the
  inflow velocities are expected to be a  fraction of the
  circular velocity \citep{Haan2009},  so it is unlikely that these radial
  velocities would be associated with nuclear inflows.

Alternatively, we can assume that the radial motions are taking place
in the plane of the host galaxy disk, as they  appear further away from the
nuclear disk as seen from the projected vertical size of the $850\,\mu$m
continuum map (see Figure~\ref{fig:ALMACO32pv_nucleardisk}).
In this case the non-circular motions
would imply a nuclear molecular gas outflow possibly due to
  molecular gas in the disk of the galaxy being swept by the AGN
  wind. This has been observed in other nearby AGN, for instance,
  NGC~1068  \citep{GarciaBurillo2014} and IC~5063 \citep{Morganti2015}.
The observations would imply an outflow on  physical scales of
70\,pc with a maximum velocity of $150-200\,{\rm km\,s}^{-1}/\sin i_{\rm
  disk}\sim 190-250\,{\rm km\,s}^{-1}$. This velocity is consistent with the
molecular outflows observed in a few Seyfert galaxies with similar AGN
bolometric luminosities \citep[see][for a compilation]{Fiore2017}, although
these are generally detected much further away from the AGN (several
hundreds of parsecs). On nuclear scales they are similar to those
identified by \cite{Davies2014} with the near-infrared H$_2$ line in
NGC~3227 and other Seyferts.

The CO(3--2) total intensity
map of the nuclear region (Figure~\ref{fig:ALMAsmallCOmaps}) shows that the emission is not
symmetric around the AGN location. Especially
along the minor axis, the kinematics reveals a cavity like
structure (Figure~\ref{fig:ALMACO32pv_nucleardisk}) which can be seen also from the CO
intensity maps and even better from the hot molecular gas map in
Figure~\ref{fig:SINFONI}.   It thus appears
that in the region
immediately southwest of the AGN the molecular gas 
and possibly the dust are being more efficiently hollowed out, although not completely emptied, 
by the AGN wind and/or jet,  as also observed in NGC~1068 \citep[see][and
  2019 in preparation]{MuellerSanchez2009, GarciaBurillo2014}. This would explain why in this region the
spectral index between Band 6 and Band 7 (see Figure~\ref{fig:ALMAspectralindex}) appears
to be mostly synchrotron-like
emission whereas the region northeast of the AGN shows evidence of both thermal and non-thermal emission.
We also note that the nuclear outflow
coincides with a region with strong evidence for the presence of shocks/outflow based on near-infrared [Fe\,{\sc ii}]
observations \citep{Schoenell2019}.

\section{Nuclear molecular gas mass and column density}\label{sec:molgascolden}

We measured the  CO(2--1) and CO(3--2) line intensities over the nuclear region
with a square aperture of $0.2\arcsec \times 0.2\arcsec$ centered at
the AGN position (see Table~\ref{tab:nuclear}). We can use the CO(2--1) measurement to
estimate the molecular gas mass in this region. 
We take a CO(1–0)/CO(2–1) brightness temperature ratio of one, 
  assuming a thermally excited and optically thick gas \citep{Braine1992}
and use the relation of \cite{Sakamoto1999} with a Galactic
CO-to-H$_2$ conversion factor $X=2 \times 10^{20}\,{\rm cm}^{-2}$ 
(K\,km\,s$^{-1}$)$^{-1}$. We obtained a molecular gas mass in the inner 15\,pc of approximately
$5\times 10^5\,M_\odot$.

 \cite{Sani2012} detected HCN emission
  using observations with an approximate $1.2\arcsec \times
  0.7\arcsec$ \, resolution. The HCN transition peaks
   at the AGN position of NGC~3227,  thus indicating that there is
   dense molecular gas. In this case, we would need to assume a brightness
  temperature ratio for the CO(1-0) and CO(2-1) transitions which is
  more appropriate for these gas conditions in the nuclear region.
  \cite{Viti2014} measured a (2--1) to (1--0) brightness temperature
  ratio of approximately 2.5 at the AGN location in NGC~1068. They
 also demonstrated that the observed suite of molecular gas transitions at
 the AGN location of NGC~1068 can be modelled with dense 
 $(>10^5\,{\rm cm}^{-3}$) and hot ($T>150\,$K) gas. If the gas conditions
 in the nuclear region of 
NGC~3227 are similar to those in NGC~1068, then the derived mass of
molecular gas (as well as the column density, see below) 
in the nuclear region  should be taken as an upper limit. This is just
due to the higher CO(2--1) to CO(1--0) ratio assumed for the calculation.

The nuclear molecular gas mass of NGC~3227  is similar to that of the
low-luminosity AGN NGC~1097 \citep{Izumi2017} but smaller than the typical values
measured by \cite{Combes2019} for low-luminosity AGN and other
Seyferts \citep{AlonsoHerrero2018, Izumi2018}. However, it agrees
remarkably well with the torus gas mass derived from SED fits of near
and mid-IR emission \citep[see][]{GarciaBernete2019}. If we
use a square aperture $0.5\arcsec$, which would encompass the nuclear
disk identified from the $850\,\mu$m continuum
(Sections~\ref{subsec:continuum_morphology} and
\ref{subsec:spectralindex}), we derived a mass in 
molecular gas of $3 \times 10^6\,M_\odot$ which agrees well with the typical values for
other nearby AGN. If we use the CO(3--2) measurement for this
0.5\arcsec \, aperture and the expressions used by \cite{Combes2019},
we would derive $7\times 10^6\,M_\odot$, which includes helium.
For reference the molecular gas mass estimates  in the central $0.8\arcsec$ are in the 
$(2-20) \times 10^7\,M_\odot$ range \citep{Davies2006, Hicks2009}, 
using the dynamical mass based on the near-infrared H$_2$ line and
assuming a gas mass fraction of 10\%. 

From the CO(2--1) measurement in the nuclear region we derive
an average H$_2$ column
density towards the AGN location (inner $\sim 15\,$pc) of approximately
$N({\rm H}_2) = 2-3 \times 10^{23}\,{\rm cm}^{-2}$ or  a factor of
  two lower if there are significant amounts of dense molecular gas in
  the nuclear region. This value would appear to be high considering that
NGC~3227 is optically classified as a Seyfert 1.5. However, X-ray, UV and
optical
studies have shown plenty of evidence for the presence of different
absorbers toward the AGN line-of-sight,
including a dusty warm absorber, lukewarm absorbers and
cold absorbers \citep[see e.g.][and references therein]{Komossa1997,
  Crenshaw2001}. Moreover, this complex X-ray absorption is also
variable \citep[see e.g.][]{Turner2018}. . Recently, \cite{Beuchert2015} derived
column densities for the ionized absorbers of $N_{\rm H} \sim
5-16\times 10^{22\,}{\rm cm}^{-2}$. They also discussed that
the absorbers could be
part of an overall clumpy medium located in the outermost dust free
broad line region or the inner part of the dusty torus.

\section{Discussion and Conclusions}\label{sec:conclusions}
\begin{figure*}
\centering
  \includegraphics[width=16cm]{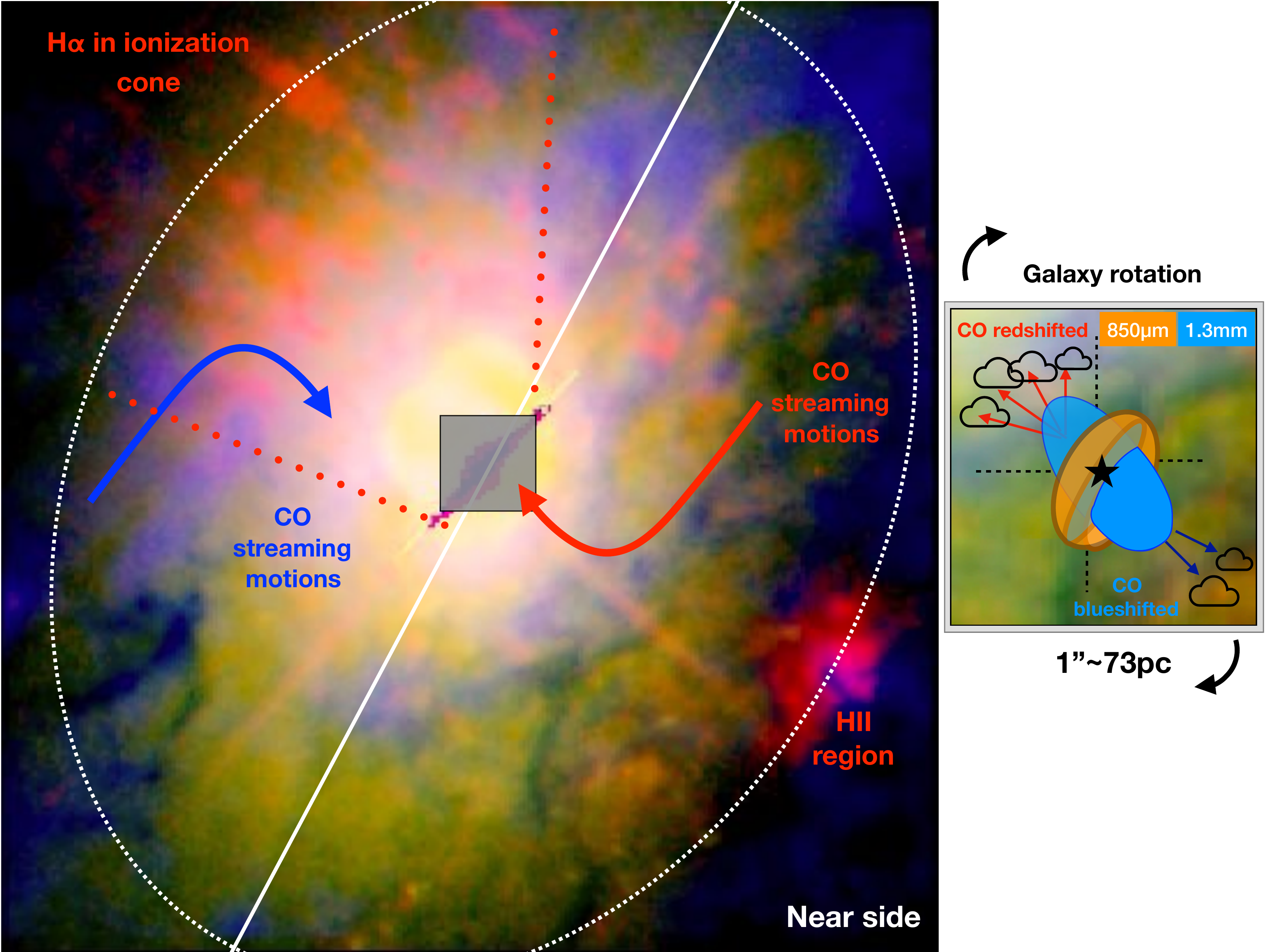}
\vspace{0.5cm}
  \caption{Summary of results for NGC~3227. The large RGB image to the
    left was
    produced using the ALMA CO(2--1) intensity in blue, the HST/ACS
    F814W image in green and the {\it HST}/ACS F658N  image in
    red. The FoV is approximately $10\arcsec \times 10\arcsec$ as in
    Figure~\ref{fig:ALMAlargeFoV}. The white straight line marks the PA of the
kinematic major axis of the    large scale disk while the dashed
magenta lines delineate the projected size of the ionization cone as
seen in the right panel of Figure~\ref{fig:ALMAlargeFoV}. The square marks
approximately the
central region with a size of $1\arcsec \times 1\arcsec$ displayed in 
Figures~\ref{fig:ALMAcontinua} and \ref{fig:ALMAsmallCOmaps}. The
small figure to the right is a cartoon (not to scale) representing the nuclear disk
mostly traced by the $850\,\mu$m continuum and the $1.3\,$mm 
extended emission perpendicular to the nuclear disk. In both panels
the arrows and their colors represent the non-circular motions
observed with the CO(2--1) and CO(3--2) transitions. 
}
             \label{fig:cartoon}%
\end{figure*}

We have presented ALMA Band 6 and Band 7 observations of the 
molecular gas using the CO(2--1) and CO(3--2) transitions and their
associated continua of the nuclear and circumnuclear regions of the
nearby Seyfert 1.5 galaxy
NGC~3227. The ALMA observations have angular resolutions
between 0.085 and 0.21\arcsec, which for this galaxy correspond to
physical resolutions in the range 7-15\,pc. With these observations we
have studied in detail the molecular gas morphology and kinematics  as well
as the (sub)millimeter continuum emission. Figure~\ref{fig:cartoon}
summarizes our results.

From a morphological point of view, the ALMA continuum
emission at 1.3\,mm and $850\,\mu$m  in the innermost $\sim 70\,$pc  shows an unresolved component ---likely to be associated with
the AGN itself--- as well as extended components. One extended component
is along the
projected direction of the ionization cone axis (${\rm PA}_{\rm cone } \sim
30-40\degr$)  and the other along the position
of the nuclear star-forming disk (${\rm PA}\sim -30\degr$) 
identified with near-infrared observations
\citep{Davies2006}. We show them in the cartoon in the right panel
of  Figure~\ref{fig:cartoon}. In the direction of the ionization cone axis the
spectral index between the two ALMA bands shows mostly positive values toward northeast of the
AGN and negative values to the southwest. For the latter we can
conclude that the ALMA continuum emission is dominated by synchrotron
emission likely associated with a putative radio jet. Along the
nuclear disk extension we derive spectral indices from slightly negative to
positive values consistent with a combination of thermal emission from
cold dust and free-emission from H\,{\sc ii} regions and non-thermal
emission due to the AGN.

Our analysis of the nuclear
radio-to-infrared SED of NGC~3227 shows that the ALMA continuum at
1.3\,mm and  $850\,\mu$m 
could be produced by a combination of different physical processes, as found for
other nearby Seyferts \citep{Krips2011, GarciaBurillo2014,
  Pasetto2019}. In particular, the extrapolation to the far-infrared of the
\cite{Nenkova2008a, Nenkova2008b} clumpy torus model fit to the
infrared emission would only account for a small fraction of the
observed $850\,\mu$m flux measured over 0.2\arcsec \, (15\,pc). 
This could be due to a significant contribution  from star formation
to the sub-mm emission in the nuclear region. 

The exquisite ALMA angular resolution reveals the distribution of the  molecular gas in the nuclear and circumnuclear regions,
with the presence of a ring-like morphology (diameter of $\sim
350\,$pc), previously
identified by \cite{Schinnerer2000}. The overall (FoV$\sim 10\arcsec
\times 10\arcsec$) kinematics of the CO(2--1) line shows 
circular motions consistent with those derived from the stellar
kinematics \citep{Davies2006, Barbosa2009}. We also detect strong streaming
motions (inflow) near the outer edges of the ring  to the southeast and northwest of
the AGN  (at projected radial distances $\sim
3-4\arcsec$) which can be explained with the presence
of a large scale bar \citep{Mulchaey1997,  Davies2014}.
These streaming motions appear to continue
inwards to regions as
close as $0.5-0.7\arcsec$ from the AGN.

In the innermost regions, the
CO(2--1) and CO(3--2) maps show a number of clumps with a morphology similar
to that of the hot molecular gas traced by the near-infrared H$_2$
line \citep{Davies2006}. None of the CO or H$_2$ peaks coincide with the AGN
location (see Figures~\ref{fig:ALMAsmallCOmaps} and \ref{fig:SINFONI}), assumed to correspond with the position of the unresolved
1.3\,mm emission. However, we also showed that the extended 1.3\,mm
and $850\,\mu$m emission in the nuclear region traces well the distribution of the
 molecular gas. The  nuclear region (inner $\sim 1\arcsec = 73\,$pc) has extremely complicated
kinematics with strong deviations from pure rotating motions along the kinematic major and minor
axes. We explored the possibility of a nuclear warp which would be 
required to explain the angle between the axis of the cone
and the normal to the galaxy plane of $\beta = 76\degr$
\citep{Fischer2013}. Although we cannot conclusively demonstrate
  the presence of a nuclear warp, a significant change of tilt between
  the host galaxy and the region collimating the ionization cone needs
  to happen in the inner 1\arcsec.


Both the CO(2--1) and the CO(3--2) transitions
show strong non-circular motions along the minor kinematic axis in the
nuclear region (Figures~\ref{fig:ALMACO21pv} and
\ref{fig:ALMACO32pv_nucleardisk}). They reach maximum velocities
of $(v-v_{\rm sys} ) \sim 190-250\,{\rm km\,s}^{-1}$ (inclination
corrected values) which are blueshifted to the southwest of the AGN
and 
redshifted to the northeast (see also the cartoon in the right panel
of 
Figure~\ref{fig:cartoon}). These motions were
well reproduced with $^{\rm 3D}$BAROLO by adding a nuclear radial velocity component to
the host galaxy rotating disk. Assuming that they are in the plane of the host
galaxy disk, then we can interpret them as due to a  nuclear molecular
outflow over scales of approximately 70\,pc.  The  molecular gas outflow velocities are
smaller than those derived from the ionized gas using the H$\alpha$ and
[S\,{\sc iii}]$\lambda$9069 emission lines, which  in this region
reach maximum values of $\pm
500\,{\rm km\,s}^{-1}$ and $\sim -900\,{\rm km\,s}^{-1}$, respectively
\citep{Fischer2013, Barbosa2009}. This is understood if the molecular
outflow in NGC~3227 is due to gas in the host galaxy entrained by the
{\it faster} AGN wind. 

Morphologically and kinematically 
the AGN wind appears to have  excavated the  molecular gas and possibly
the dust in regions close to the AGN. This created a cavity-like
structure, which is more apparent to the southwest of the AGN 
(see Figure~\ref{fig:ALMAsmallCOmaps} and especially Figure~\ref{fig:SINFONI})
that is similar to that observed in NGC~1068 \cite[see][and
  2019]{MuellerSanchez2009, GarciaBurillo2014}. We can estimate the
nuclear  outflow rate on both sides of the AGN along the minor
axis taking expression given in equation~4
of \cite{GarciaBurillo2014}: 
$dM/dt = 3 \times v_{\rm out} \times M_{\rm mol}/(R_{\rm out} \times
\tan(i_{\rm disk}))$ for a multi-conical outflow
uniformly filled by the outflowing clouds. We  measured the CO(2-1)
fluxes in square apertures of 0.2\arcsec \, on a side just outside of the central
AGN position and estimate molecular gas masses (following the same
assumptions as in Section~\ref{sec:molgascolden}) of $7\times 10^5\,M_\odot$ and $8\times 10^4\,
M_\odot$ to the northeast and southwest of the AGN, respectively. For
the outflow velocity we take a typical value of $v_{\rm out} =
100\,{\rm km\,s}^{-1}$ and a size for the outflow of $R_{\rm out } =
35\,$pc. This results in nuclear molecular outflow rates of $dM/dt = 5\,M_\odot\,{\rm
  yr}^{-1}$ and $dM/dt = 0.6\,M_\odot\,{\rm
  yr}^{-1}$ to the northeast and southwest of the AGN. These values
are about an order of magnitude lower than the molecular outflow rate
at the circumnuclear disk of NGC~1068 \citep{GarciaBurillo2014} but
comparable to that measured for NGC~1433 \citep{Combes2013}.

At the AGN
position and with our angular resolution, this region is not entirely devoid of
molecular gas. We estimated  a mass in molecular gas of 
$5\times 10^5\,M_\odot$ (for a square aperture 0.2\arcsec $\sim$15\,pc on a side) and
an equivalent average column density of $N({\rm H}_2) = 2-3 \times 10^{23}\,{\rm
  cm}^{-2}$ at the AGN position. However, if the molecular gas
  at the AGN location is dense and relatively hot as in NGC~1068, the
  derived molecular gas mass and column densities would be
  approximately a factor of two lower. The lower values are   due to the higher
  CO(2--1) to CO(1--0) ratio observed in the nuclear region of NGC~1068
  compared to the typical ratios observed in the disks of galaxies. Nevertheless,  although
  NGC~3227 is classified as a Sy1.5,   there is evidence of a gas-rich nuclear/circumnuclear
environment. This agrees with the weak relationship between the
circumnuclear ($1-3\,$kpc) and nuclear (tens of parsecs)
molecular gas mass and the  X-ray column density in nearby AGN
\citep[see][respectively]{Rosario2018, Combes2019}
and implies that the molecular gas has a low filling factor. As a
quick estimation, the gas mass and size of the region give an average
filling factor of molecular gas, assuming a density of $\sim 10^6\,$cm$^{-3}$,
of $6\times 10^{-3}$, or a $\sim 80$\% chance that a given line-of-sight
will not intersect
a molecular clump. In this sense, it is not surprising that the AGN is
partially unobscured. 

In summary, the nuclear CO(2--1) , CO(3--2)   and dust morphology of
NGC~3227 do 
not resemble a {\it classical} compact torus. Rather, the dust and molecular gas 
emission is in a nuclear disk extended over several tens of parsecs. The orientation of the
  nuclear disk is consistent with being part of the collimating
  structure of the ionization cone. We find evidence that the dust might be heated not only by the
  AGN but also by the nuclear on-going/recent star formation activity
  \citep{Davies2006, Esquej2014}. In the CO(2--1) and CO(3--2) transitions we do
  not observe a bright and isolated nuclear structure/torus, but several
  clumps well connected with the circumnuclear ring in the host galaxy
  disk. In this situation it is
  difficult to say where the torus {\it starts}, although such 
  molecular gas morphologies are also common in other  AGN
 of similar or lower luminosity  \citep{Izumi2017, Izumi2018, AlonsoHerrero2018, Combes2019}.


\begin{acknowledgements}
We are grateful to an anonymous referee for constructive comments
which helped improve the manuscript.
We thank Luis Colina, Alvaro Labiano, Leonard Burtscher and Daniel Asmus for useful discussions.
A.A.-H. and I.G.-B. acknowledge support from the Spanish Ministry
of Science, Innovation and Universities through grant
AYA2015-64346-C2-1-P, which was party funded by the FEDER 
program. A.A.-H. also acknowledges  CSIC grant
PIE201650E36. A.A.-H., S.G.-B. and A.U. acknowledge support from the Spanish Ministry
of Science, Innovation and Universities through grant
PGC2018-094671-B-I00, which was party funded by the FEDER 
program. 
S.G.-B. and C.R.A. acknowledge the Spanish Plan Nacional de Astronom\' ia
y Astrof\' isica under grant AYA2016-76682-C3-2-P. 
M.P.-S. and D.R. acknowledge support from STFC through grant ST/N000919/1.
T.D.-S. acknowledges support from ALMA-CONICYT project 31130005 and
FONDECYT regular project 1151239. 
S.F.H. is supported by European Research Council Starting Grant ERC-StG-677117 DUST-IN-THE-WIND.
C.R.A. also acknowledges the Ram\'on y Cajal Program of the Spanish
Ministry of Science and Technology through project RYC-2014-15779. 
S.R. and M.V. gratefully acknowledge support from the Independent
Research Fund Denmark via grant numbers DFF 4002-00275 and
8021-00130. L.K.H. acknowledges funding from the INAF PRIN-SKA 2017
program 1.05.01.88.04. C.R. acknowledges support from the CONICYT+PAI
Convocatoria Nacional subvencion a instalacion en la academia
convocatoria a\~{n}o 2017 PAI77170080.  D.R. acknowledges the support
of the Science and Technology Facilities Council (STFC) through grant
ST/P000541/1. AU acknowledges support from the
Spanish MINECO grants ESP2015-68964 and AYA2016-79006.

This paper makes use of the following ALMA data:
ADS/JAO.ALMA.2016.1.00254.S and JAO.ALMA.2016.1.01236.S. ALMA is a partnership
of ESO (representing its member states), NSF (USA) and NINS
(Japan), together with NRC (Canada), MOST and ASIAA
(Taiwan), and KASI (Republic of Korea), in cooperation with
the Republic of Chile. The Joint ALMA Observatory is
operated by ESO, AUI/NRAO and NAOJ. Based on observations made with
the NASA/ESA Hubble Space Telescope, obtained from the data archive at
the Space Telescope Science Institute. STScI is operated by the
Association of Universities for Research in Astronomy, Inc. under NASA
contract NAS 5-26555. 
\end{acknowledgements}

%
%

   \bibliographystyle{aa} 
   \bibliography{bibliography} 

\end{document}